\documentclass{aa}  
\usepackage{txfonts}
\usepackage{graphicx}
\usepackage[breaklinks=true, colorlinks=true, citecolor=blue, linkcolor=blue,urlcolor=blue]{hyperref}
\usepackage{multicol}
\usepackage{orcidlink}
\usepackage{comment}
\usepackage{enumitem}

\newcommand{\vir}[1]{``#1''}
\renewcommand{\eqref}[1]{equation\ (\ref{#1})}
\newcommand{\bba}{$^{\scriptstyle 3\mathrm{D}}$B{\sc arolo}}
\newcommand{\hi}{\ifmmode{\rm HI}\else{H\/{\sc i}}\fi}
\newcommand{\de}{\ifmmode{^\circ}\else{$^\circ$}\fi} 
\newcommand{\vlrs}{\ifmmode{V_\mathrm{LSR}}\else{$V_\mathrm{LSR}$}\fi}
\newcommand{\mo}{{\rm M}_\odot}
\newcommand{\moyr}{{\rm M}_\odot \, {\rm yr^{-1}}}

\newcommand{\kms} {\,{\rm km\,s}^{-1}}

\newcommand{\cloudnum}{19}
\newcommand{\co}{$^{12}$CO(2$\rightarrow$1)} 
\newcommand{\cof}{$^{12}$CO(4$\rightarrow$3)}
\newcommand{\rms}{\ifmmode{\sigma_{\rm rms}}\else{$\sigma_{\rm rms}$}\fi} 
\newcommand{\xcoun}{\mathrm{cm^{-2} \, (K\, \kms)^{-1} }}
\newcommand{\acoun}{\mathrm{M_\odot \, (K\, \kms\, pc^2)^{-1} }}
\newcommand{\mmol}{M_\mathrm{mol}}
\newcommand{\mvir}{M_\mathrm{vir}}
\newcommand{\avir}{\alpha_\mathrm{vir}}
\newcommand{\sigv}{\sigma_\mathrm{v}}
\newcommand{\fmol}{f_\mathrm{mol}}

\defcitealias{DiT+18}{DT18}
\defcitealias{Miville-Deschenes+2017}{MD17}

\begin{document} 

\title{A survey of molecular clouds in the Galactic center's outflow}

\author{
Enrico M.\ Di Teodoro\inst{1,2}\orcidlink{0000-0003-4019-0673}
\and Mark Heyer\inst{3}\orcidlink{0000-0002-3871-010X}
\and Mark R.~Krumholz\inst{4}\orcidlink{0000-0003-3893-854X}
\and Lucia Armillotta\inst{1,2,5}\orcidlink{0000-0002-5708-1927}
\and \\
Felix J.\ Lockman\inst{6}\orcidlink{0000-0002-6050-2008}
\and
Andrea Afruni\inst{1}\orcidlink{0000-0002-2858-6950} 
\and Michael P. Busch\inst{7}\orcidlink{0000-0003-4961-6511}
\and N. M. McClure-Griffiths\inst{4}\orcidlink{0000-0003-2730-957X} 
\and \\
Karlie~A.~Noon\inst{4}\orcidlink{0000-0002-9699-6863}
\and Nicolas Peschken\inst{1}\orcidlink{0000-0002-9925-2974}
\and Qingzheng Yu\inst{1}\orcidlink{0000-0003-3230-3981}
}

\institute{
Universit\`{a} di Firenze, Dipartimento di Fisica e Astronomia, via G. Sansone 1, 50019 Sesto Fiorentino, Firenze, Italy\\\email{enrico.diteodoro@unifi.it}
\and 
INAF - Osservatorio Astrofisico di Arcetri, Largo E. Fermi 5, I-50125 Firenze, Italy
\and 
Astronomy Department, University of Massachusetts, Amherst, MA 01003, USA
\and 
Research School of Astronomy and Astrophysics - The Australian National University, Canberra, ACT, 2611, Australia
\and 
Department of Astrophysical Sciences, Princeton University, Princeton, NJ 08544, USA
\and 
Green Bank Observatory, Green Bank, WV 24944, USA
\and 
National Radio Astronomy Observatory, 520 Edgemont Road, Charlottesville, VA 22903, USA
}

\abstract{
    The nucleus of the Milky Way is known to drive a large-scale, multiphase galactic outflow, with gas phases ranging from the hot highly-ionized to the cold molecular component.
    In this work, we present the first systematic search for molecules in the Milky Way wind. 
    We use the Atacama Pathfinder EXperiment (APEX) to observe the \co\ emission line in \cloudnum\ fields centered on previously known high-velocity atomic hydrogen (\hi) clouds associated with the outflow. 
    Over 200 CO clumps are detected within 16 different \hi\ clouds. 
    These clumps have typical radii of $1-3$ parsec, high velocity dispersions of $1-6 \, \kms$ and molecular gas masses ranging from a few to several hundreds $\mo$.
    Molecular clumps in the wind sit on the low-mass end of the mass - size relation of regular molecular clouds, but are far displaced from  the mass (or size) - linewidth relation, being generally more turbulent and showing high internal pressures. 
    Nearly $90\%$ of the clumps are gravitationally unbound with virial parameters $\avir \gg 10 - 100$, 
    indicating that these structures are either being disrupted or they must be confined by external pressure from the surrounding hot medium.
    While the observed properties of CO clumps do not seem to evolve clearly with latitude, we find that molecular gas is not detected in any of the 6 \hi\ clouds with projected distances over 1 kpc from the Galactic Center, suggesting the existence of a maximum timescale of $\sim3$ Myr for the dissociation of molecular gas within the wind. 
    Overall, current observations in the Galactic center support a scenario in which a hot wind entrains cold gas clouds from the disk, driving their progressive transformation from molecular to atomic and ultimately ionized gas through stripping, turbulence, and dissociation.
}


\maketitle

\section{Introduction}
Our Galaxy, the Milky Way (MW), hosts a large-scale, multiphase gas outflow in its nuclear regions \citep[see review by][]{Sarkar2024}. 
The $\gamma$-ray Fermi bubbles \citep{Su+10} and the X-ray eROSITA bubbles \citep{Predehl+2020} are striking evidence of fast-moving outflowing gas containing both hot ($T>10^6$ K) and non-thermal (particle energy $\gtrsim$ GeV) components, extending to over 10 kpc from the Galactic Center (GC).
Ultra-violet (UV) absorption studies have revealed the presence of high-velocity warm ionized gas ($T\sim10^{4-5}$ K) across the volume of these bubbles \citep{Fox+15,Bordoloi+17,Cashman+2023}.
Closer to the Galactic plane, a population of high-velocity cool neutral gas ($T\sim10^{3-4}$ K) clouds associated with the outflow were also detected thanks to observations of the atomic hydrogen (\hi) emission line at 21 cm \citep{McClure-Griffiths+13,DiT+18}.
Some of these clouds host cold molecular gas cores ($T\sim 10^{1-3} K)$, which have been detected either in carbon monoxide (CO) emission \citep{DiTeodoro+2020,Heyer+2025} or in H$_2$ absorption \citep{Cashman+2021}.
These outflows could be powered by either a recent activity from Sagittarius A$^*$ \citep[e.g.,][]{Zubovas+2012,Bland-Hawthorn+2019} or by the intense star formation in the inner Galaxy \citep[e.g.,][]{Crocker+2015,Sarkar+2015}, both in the Central Molecular Zone \citep[CMZ, e.g.,][]{Barnes+2017, Henshaw+2023} and in the nuclear star cluster \citep[e.g.,][]{Schodel+2014, Nogueras-Lara+2020}.

Similar multiphase outflows, driven by either intense star formation or active galactic nucleus (AGN) activity, are observed in many galaxies across the Universe \citep[e.g.][]{Veilleux+20}. 
These phenomena play a crucial role in the evolution of galaxies. By circulating and removing metal-enriched gas, they regulate star formation and stellar mass assembly, as well as the distribution of metals within and around galaxies \citep[e.g.,][]{Naab&Ostriker17}.
Over the years, numerous theoretical studies have employed high-resolution (magneto)hydrodynamical (MHD) simulations to investigate the detailed physics underlying the launching and the evolution of galactic winds \citep[see review by][]{Thompson+2024}.
A general consensus from these studies is that, while the hot gas phase carries most of the energy, a significant fraction of the wind's mass resides in the cool and cold phases \citep[e.g.,][]{Kim+20, Schneider+20,Wibking+23, Vijayan24a, Vijayan25a}, especially at the wind base.
It is therefore particularly important to understand the role of cooler phases in galactic winds.

The origin and fate of cool/cold clouds within a hot wind pose significant physical challenges.
Clouds lifted from the disk and entrained within the hot wind must be accelerated to high velocities through one or more mechanisms, including gain of momentum from the hot gas \citep[e.g.,][]{Schneider+20,Fielding+2022}, radiation pressure \citep[e.g.,][]{Thompson+15,Zhang+2018} and cosmic ray pressure \citep[e.g.,][]{Peschken+23, Rathjen+23, Armillotta+2024}.
However, the interaction between gas phases is typically destructive, and cold clouds are expected to be shredded and disrupted on short timescales \citep[e.g.,][]{ Scannapieco&Bruggen15,Schneider+17}.
Radiative cooling of the hot fast-moving gas, induced by turbulent mixing with a seed cloud, have been invoked to explain the cool gas survival \citep[e.g.,][]{Gronke+20, Banda-Barragan+21, Kanjilal+2021}.
In this case, the cloud gains mass while also acquiring momentum from the faster hot gas.
Other physical mechanisms that may extend cloud lifetimes include magnetic fields \citep[e.g.,][]{Sparre+20, Gronnow+22}, and, in some cases, thermal conduction \citep[e.g.,][]{Bruggen&Scannapieco16, Armillotta+17}.
An alternative scenario proposes that cold gas can be formed out of the hot phase through thermal instabilities, once the wind has cooled enough by adiabatic expansion at large distances \citep[e.g.,][]{Thompson+16b,Richings+18}.
Despite all these possible scenarios, it remains uncertain how effectively these processes operate under realistic conditions. 
High-resolution observations of cool gas in galactic winds are therefore fundamental to constrain these theoretical models of cloud acceleration and survival. 

The nucleus of the MW is one of the few places in the Universe where we can observe individual cool clouds in an outflow and investigate their resolved structure, physical state, and kinematics on (sub)-parsec scales.
A pioneering study by \citet{McClure-Griffiths+13} first identified a population of 86 high-velocity \hi\ clouds in the inner Galaxy ($|\ell|<5^\circ, |b|<5^\circ)$, observed with the Australia Telescope Compact Array (ATCA), and proposed that they may represent the cool atomic counterpart of the nuclear outflow. 
Follow-up deep surveys with the Green Bank Telescope (GBT) supported this interpretation, expanding the known distribution of \hi\ clouds to $\pm{15\de}$ ($\simeq 2$ kpc) in both Galactic longitude and latitude \citep{DiT+18}. 
Recent data show that this population may extend even farther, reaching $b\sim30\de$ \citep{Bordoloi+2025}.
Identified \hi\ clouds show anomalous local standard of rest (LSR) velocity, i.e., their kinematics are not compatible with the rotation of the Galactic disk but rather with a bi-conical wind accelerating from the GC and reaching velocities of $\sim300-350$ $\kms$ \citep{Lockman+20}.
These data provided a first direct estimate of the acceleration/velocity of the MW’s cool wind, and gave some important clues on other key parameters of the outflow, including its geometry, cool mass loading rate, luminosity and lifetime.

Targeted follow-up observations with the Atacama Pathfinder EXperiment (APEX) of the \co\ emission line in two outflowing \hi\ clouds (MW-C1 and MW-C2) revealed the presence of dense molecular clumps \citep{DiTeodoro+2020}. 
More recently, $^{12}$CO(1$\rightarrow$0) emission from a third cloud (MW-C21), detected with the Large Millimiter Telescope (LMT), further expanded the inventory of known molecular clouds associated with the nuclear wind \citep{Heyer+2025}. 
These pilot CO observations suggest that outflowing \hi\ clouds may carry substantial amounts of molecular matter, up to $60\%-80\%$ of their total mass \citep{DiTeodoro+2020}.
They also provide some important insights into the physical state of the molecular clouds: they appear to be too molecule-rich to be in chemical equilibrium \citep{Noon+2023}, dominated by compressive turbulence \citep{Gerrard+2024}, and gravitationally unbound \citep{Heyer+2025}.
Overall, the evidence gathered so far, though based on only three molecular clouds, points toward a scenario in which molecular gas is swept up from the Galactic disk and entrained into the hot wind, rather than reformed \textit{in situ}.
In this picture, molecular gas is expected to gradually dissociate under the influence of the Galactic extreme ultra-violet radiation field \citep{Vijayan24b}, and the clouds should also be shredded as they are accelerated away from the Galaxy.
Additional support for a disk origin comes from observations of molecular gas at lower latitudes ($|b|\lesssim3^\circ$), likely entrained by the nuclear wind.
In particular, CO emission has been detected in molecular structures in correspondence of the X-ray Galactic chimneys \citep{Ponti+2019,Ponti+2021}, up to 100 pc from the GC \citep{Veena+2023}, and, on much larger scales, at the disk-halo interface near the edge of the MW's inner \hi\ cavity \citep{Lockman+1984,Lockman+16}, at Galactocentric radii of $R\simeq 3$ kpc \citep{Su+2022}.
However, low-latitude ($|b|<1\de$) features are hard to attribute uniquely to the wind, as anomalous velocity structures can also be produced
by non-circular streaming motions and shocks along the Galactic bar \citep[e.g.,][]{Sormani+2018}.

Despite these important advances, a comprehensive characterization of the physical state of the cold gas entrained in the wind, its interaction with other gas phases, and the evolution and fate of these clouds remains elusive.
One of the main limitations so far has been the very small number of molecular clouds observed.
Our current understanding relies on only three CO detections, which precludes any statistical analysis.
This work is a first step toward addressing this issue.
Here we present the results of the first systematic exploration of the molecular gas content in the MW's nuclear wind.
We expand the sample of clouds observed in \co\ with APEX, adding 17 new objects to the 2 clouds presented in \citet{DiTeodoro+2020}.
Their physical and kinematic properties are investigated and compared with regular molecular clouds in the disk.

The remainder of this article is structured as follows. Section~\ref{sec:data} presents our new observations of molecular gas in the GC outflow. 
In Section~\ref{sec:methods} and Section~\ref{sec:radmod}, we describe in detail the techniques used for the analysis of the data and to derive the physical properties of the outflowing gas. 
We present our results on the molecular clump properties, scaling relations and evolution in Section~\ref{sec:results}, discussing their implications in Section~\ref{sec:discussion}. 
Finally, we summarize and conclude in Section~\ref{sec:conclusion}.
Throughout this paper, we assume a distance to the GC of $D=8.2$ kpc \citep{Gravity19}, for which 25 arcsec correspond to about 1 pc.

\section{Data}
\label{sec:data}

\subsection{Cloud sample}
\begin{figure}
    \centering
    \includegraphics[width=0.48\textwidth]{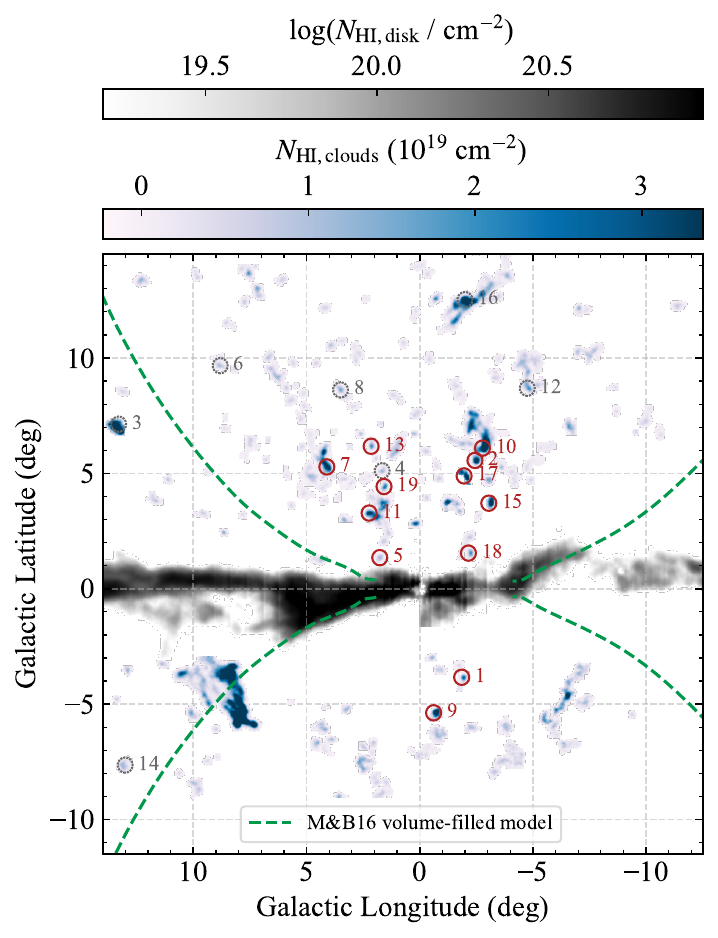}
    \caption{The sample of high-velocity clouds studied in this work. The grey-scale map represents the MW disk \citep{Lockman+16}, while the blue colorscale map the outflowing \hi\ high-velocity clouds \citep{DiT+18}.
    Red circles mark the clouds followed-up in the \co\ and \cof\ lines with APEX and analysed in this work, numbered as in \autoref{tab:sample}. Dotted gray circles denote clouds not detected in CO emission.}
    \label{fig:sample}
\end{figure}

To date, over 200 outflowing \hi\ clouds have been detected in the inner 15 degrees of our Galaxy, including the 86 clouds published in \citet{McClure-Griffiths+13} and the 106 clouds cataloged in \citet{DiT+18}. 
\autoref{fig:sample} shows a map of these high-velocity clouds.
The grayscale map represents the \hi\ emission at the tangent points from the Galactic All Sky Survey \citep[GASS,][]{McClure-Griffiths+09}, roughly tracing the gaseous disk in the inner region of the MW \citep{Lockman+16}. 
The bluescale colormap shows instead the \hi\ column densities of outflowing gas clouds from our GBT survey of the GC, including both data from \citet{DiT+18} and new data from more recent observations (Yu et al., in prep., Lockman et al., in prep.). 
As a reference, we also show the volume-filled model for the Fermi bubbles by \citet{Miller&Bregman16} as green dashed lines, representing the alleged boundaries of the bubbles.
All these clouds have anomalous kinematics and are not in Galactic rotation.
Most of them are very compact and have relatively low column densities (up to a few $10^{19}$ cm$^{-2}$), with the exception of a few larger and denser structures, like the well-known \vir{Shane feature} at $(l,b) \simeq (8\de,-5\de)$ \citep[e.g.,][]{Burton+1978}.

In this work, we investigate the molecular gas component in a subsample of \cloudnum\ outflowing \hi\ clouds \citep[including MW-C1 and MW-C2 from][]{DiTeodoro+2020}, selected from a new updated catalog that combines the catalogs published in \citet{McClure-Griffiths+13} and \citet{DiT+18} with new detections found in our latest GBT \hi\ observations. 
The clouds targeted are shown with red or gray circles in \autoref{fig:sample} and listed in \autoref{tab:sample}.
This sample was chosen to cover a wide range in \hi\ column densities and in Galactic latitude/longitude, with the idea of investigating both the relation between \hi\ and molecular gas and the evolution of the cold gas as a function of distance from the GC.
Out of \cloudnum\ clouds, three targets lie below and sixteen above the Galactic plane. 
This North/South asymmetry mirrors that of the parent \hi\ population and arises primarily from observational biases in our \hi\ surveys (e.g., reduced GBT sensitivity and coverage at $b<0\de$) rather than from any intrinsic asymmetry in the wind \citep[see][]{DiT+18}.
Peak \hi\ column densities of the APEX targets measured from the GBT data range between $5\times10^{18}$ cm$^{-2}$ and $8\times10^{19}$ cm$^{-2}$. 
We emphasize that the GBT has a relatively low spatial resolution $($570''$)$ and that the measured column densities may be several times lower than the real ones because of beam dilution \citep[see also][]{Noon+2023}.
For an assumed distance from the GC of 8.2 kpc, these clouds have \hi\ masses between a few tens and a few thousand solar masses. 
The LSR velocities of the targets span from -260 $\kms$ to +370 $\kms$, while typical \hi\ line widths are in the range $5-30$ $\kms$ (Full Width at Half Maximum, FWHM). 

{
\setlength{\tabcolsep}{5pt} 
\begin{table}
\centering
\caption{Observational properties of the cloud sample observed with APEX.}
\label{tab:sample}
\begin{tabular}{lccccc}
\noalign{\vspace{5pt}}\hline\hline\noalign{\vspace{5pt}}
Name & $\ell$ & $b$ & $\vlrs$ & OTF & $\sigma_\mathrm{rms}$ \vspace{2pt}\\
   & ($\de$) & ($\de$) & ($\kms$) & & (mK) \\
\noalign{\vspace{5pt}}\hline\hline\noalign{\vspace{5pt}}
MW-C1 & $-1.85$ & $-3.83$ & $+163$ & $15'\times 15'$ & 49 \\
MW-C2 & $-2.44$ & $+5.56$ & $+265$ & $15'\times 15'$ & 34 \\
MW-C3$^\dagger$ & $+13.30$ & $+7.13$ & $-100$ & $15'\times 15'$ & 65 \\
MW-C4$^\dagger$ & $+1.66$ & $+5.11$ & $-262$ & $10'\times 10'$ & 80 \\
MW-C5 & $+1.77$ & $+1.35$ & $-178$ & $15'\times 15'$ & 56 \\
MW-C6$^\dagger$ & $+8.82$ & $+9.67$ & $+200$ & $10'\times 10'$ & 43 \\
MW-C7$^\ddagger$ & $+4.10$ & $+5.28$ & $+205$ & $10'\times 10'$ & 51 \\
MW-C8$^\dagger$ & $+3.50$ & $+8.62$ & $+365$ & $10'\times 10'$ & 55 \\
MW-C9$^\ddagger$ & $-0.61$ & $-5.37$ & $-175$ & $15'\times 15'$ & 53 \\
MW-C10$^\ddagger$ & $-2.77$ & $+6.10$ & $-165$ & $15'\times 15'$ & 55 \\
MW-C11$^\ddagger$ & $+2.24$ & $+3.28$ & $-203$ & $15'\times 15'$ & 55 \\
MW-C12$^\dagger$ & $-4.75$ & $+8.69$ & $-188$ & $10'\times 10'$ & 51 \\
MW-C13 & $+2.15$ & $+6.18$ & $-217$ & $15'\times 15'$ & 60 \\
MW-C14$^\dagger$ & $+13.01$ & $-7.64$ & $-182$ & $15'\times 15'$ & 50 \\
MW-C15 & $-3.05$ & $+3.71$ & $+168$ & $15'\times 15'$ & 65 \\
MW-C16$^\dagger$ & $-1.99$ & $+12.53$ & $-136$ & $15'\times 15'$ & 70 \\
MW-C17 & $-1.96$ & $+4.89$ & $+230$ & $15'\times 15'$ & 57 \\
MW-C18 & $-2.15$ & $+1.55$ & $-258$ & $15'\times 15'$ & 57 \\
MW-C19 & $+1.58$ & $+4.43$ & $-226$ & $15'\times 15'$ & 59 \\
\noalign{\vspace{3pt}}\hline \noalign{\vspace*{3pt}}
\multicolumn{2}{l}{$^\dagger$ Not detected.} &
\multicolumn{3}{l}{$^\ddagger$ Multi-velocity detections.}
\end{tabular}
\tablefoot{
Column 1 lists our assumed name, columns 2-4 give the Galactic coordinates and central velocity of the target, column 5 is the size of the on-the-fly map, and column 6 is the rms noise of the datacubes at 230 GHz in a 0.5 $\kms$ channel.
}

\end{table}
}

\subsection{Observations and data reduction}
Millimeter and sub-millimiter data presented in this work were collected with the Atacama Pathfinder EXperiment (APEX) telescope under European Southern Observatory (ESO) dedicated time during observing periods P104 and P106 (Proposal IDs: 0104.B-0106A, 0106.B-0034A and 0106.C-0031A, Principal Investigator: Di Teodoro). 
Observations with the 12 m APEX antenna were spread across several ESO sessions, from November 2019 to April 2022. 
The weather conditions varied across the different observing sessions, with a precipitable water vapour (PWV) that ranged approximately between 0.5 mm and 3.5 mm.

Observations of MW-C1 and MW-C2 for the P104 period were carried out with the PI230 receiver tuned to map the \co\ emission line at 230.538 GHz. 
These data have been described and analysed in \citet{DiTeodoro+2020}.
For the P106 period, we targeted 17 additional clouds in the \co\ line with the new nFLASH230 receiver, which replaced PI230 in 2020.
Exploiting the possibility of using in parallel the nFLASH460 receiver, we 
also tried to detect the \cof\ emission line at 461.041 GHz at the same time.
The two spectrometer processors unit (FFTS) backends provide a bandwidth of 8 GHz for nFLASH230 and 4 GHz for nFLASH460, with a spectral resolution of 61 kHz, corresponding to roughly 0.08 $\kms$ at 230 GHz and to 0.04 $\kms$ at 460 GHz. 
The FWHM beam sizes are $\theta_{230} = 28.6''$  and $\theta_{460} = 14.3''$ at 230 GHz and at 460 GHz, respectively. 
These values correspond to linear resolutions of 1.1 pc and 0.6 pc at the distance of the GC.
Average main beam efficiencies during the period of observations were $\eta_\mathrm{mb}= 0.80 \pm 0.06$ for nFLASH230 and $\eta_\mathrm{mb}= 0.60 \pm 0.05$ for nFLASH460\footnote{https://www.apex-telescope.org/telescope/efficiency/}.
Targets were observed in position-switching, on-the-fly (OTF) mapping mode, with sub-map sizes of $5'\times5'$, sampling every $9''$, and using a dumping time of 1 second. 
For each target, either four or nine contiguous $5'\times5'$ submaps were covered, in order to reach a final field size of either $15'\times15'$ or $10'\times10'$.
The coordinate centers of each map were chosen to align with the peak of the column density of \hi\ derived from the GBT data.
Integration times for each target were roughly 20 hours for the $15'\times15'$ maps and 10 hours for the $10'\times10'$.
 Map centers and sizes for all clouds in our sample are listed in \autoref{tab:sample}.

The data reduction was performed using the Continuum and Line Analysis Single-dish Software \citep[CLASS,][]{Gildas+13} and following standard data reduction steps. 
The data provided by APEX are spectra calibrated in antenna temperature scale $T^*_A$, which were converted to main beam brightness temperature through the measured main beam efficiency, i.e.\ $T_\mathrm{mb} = T^*_A/\eta_\mathrm{mb}$.
A first-order baseline was then subtracted from each calibrated spectrum. During the process, we masked velocity channels where we expected to see emission based on the \hi\ data, both from the high-velocity clouds and from the local interstellar medium at near-zero LSR velocity.
Finally, after smoothing and resampling the spectra to a channel width of 0.5 $\kms$ in the spectral range $-400\, \kms \leq V_\mathrm{LSR} \leq +400\, \kms$ , we gridded them into datacubes with a pixel size of $9''$ at 230 GHz and of $5''$ at 460 GHz, approximately one third of the expected main beam sizes.
The root mean square (rms) noise $\sigma_\mathrm{rms}$ in our final datacubes at 230 GHz ranges between 34 and 80 mK for a 0.5 $\kms$ channel (see \autoref{tab:sample}), while at 460 GHz the rms noise varies between 200 mK and 500 mK.
As expected, none of the targets were detected in the \cof\ emission line, thus in the rest of this work we will only focus on the analysis of the \co\ line data.

Finally, 
we also report on a failed attempt to detect the HCN(2$\rightarrow$1) emission line at 177.239 GHz and the HCO$^+$(2$\rightarrow$1) emission line at 178.376 GHz in the two densest CO clumps observed in MW-C1 \citep[see][]{DiTeodoro+2020}. 
HCN and HCO$^+$ lines typically trace denser molecular gas than \co, with volume densities up to $n \simeq 10^4$ cm$^{-3}$.
We used single pointing observations with the SEPIA180 receiver on APEX, integrated for about 3 hours for each clump, and reached rms noises in the reduced spectra of $7-10$ mK for a 0.5 $\kms$ channel width. 
Neither the HCN nor the HCO$^+$ species were detected in these two outflowing molecular clumps.

\section{Cloud identification and characterization}
\label{sec:methods}
For the analysis of the \co\ datacubes for our \cloudnum\ targets, two main codes were exploited: \bba\ \citep[3DB, hereinafter,][]{DiTeodoro+2015} and \textsc{pycprops} \citep{Rosolowsky+2021}.
The former is a code for galaxy kinematics, but it also features several useful functions to analyse emission line datacubes. 
In particular, 3DB was used to build masks that identify regions of genuine line emission and to derive moment maps from the datacubes. 
\textsc{Pycprops} is an improved, \textsc{python} implementation of the \textsc{cprops} algorithm \citep{Rosolowsky+2006} that was developed to study the properties of resolved molecular clouds in the PHANGS-ALMA galaxy survey \citep{Leroy+2021}. 
This code can be used to assign voxels with emission to different sources and to calculate physical quantities, such as cloud position and size, integrated flux, luminosity and molecular gas mass.

\subsection{Emission line detection, masking and moment maps}
Regions of line emission in the datacubes are identified through the 3D source finder implemented in 3DB. 
In short, this algorithm identifies regions of contiguous voxels that lie above some primary flux threshold and then grows them adding more voxels at their edges down to a lower secondary threshold. 
Detections are then merged together based on criteria of proximity in both the spatial and spectral domains and finally they can be rejected/accepted depending on their size and/or spectral extent.

This approach relies on the noise level being constant throughout the datacube.
Because each of our final datacubes is made up of several OTF sub-maps acquired during various observational runs and under different weather conditions, the noise level generally is not constant, but it can vary spatially from sub-map to sub-map. 
To overcome this issue, we run the source finder on signal-to-noise (S/N) cubes rather than on the original data. 
To create the S/N cubes, we first derive noise maps $\sigma(x,y)$ for each target by calculating the noise level for each spectrum in each spatial pixel. 
The noise level is assumed as the standard deviation of the data, computed via robust statistics, i.e.\ $\sigma(x,y) = 1.4826\,\mathrm{MADFM}(s(x,y))$, where MADFM is the median absolute deviation from the median of the spectrum $s(x,y)$ at the spatial position $(x,y)$ and the factor 1.4826 converts between MADFM and standard deviation for normally distributed data.
Finally, each channel map of the original cubes is divided by the noise maps to obtain the S/N cubes.

Source finding is performed on the resulting S/N cubes in the full LSR spectral range $\pm400\,\kms$, using a primary threshold of 3.5 and a secondary threshold of 2.
We require reliable detections to cover at least the size of the beam and to extend over 1 $\kms$ ($ = 2$ channels).
The source finder typically detects CO emission from local molecular gas in the Solar neighborhood at $\vlrs\sim0\, \kms$ plus CO emission from high-velocity gas clouds. 
Line emission from local gas is detected in all observed fields but MW-C6.
Because we are interested in studying only high-velocity clouds that may be associated with the GC outflow, we discard all detections with a velocity that deviates less than 50 $\kms$ from the expected Galactic rotation. 
High-velocity molecular gas is detected in 9 out of the \cloudnum\ observed fields. 
Datacubes with no clear detections were smoothed to a channel width of 1 $\kms$ and 2 $\kms$ to reduce the noise level and the source finding process was repeated.
This allows us to detect faint emission in 3 additional fields. 
All cubes were also carefully inspected to make sure that the source finder did not miss any faint but convincing emission. 

\begin{figure*}
    \centering
    \includegraphics[width=0.87\textwidth]{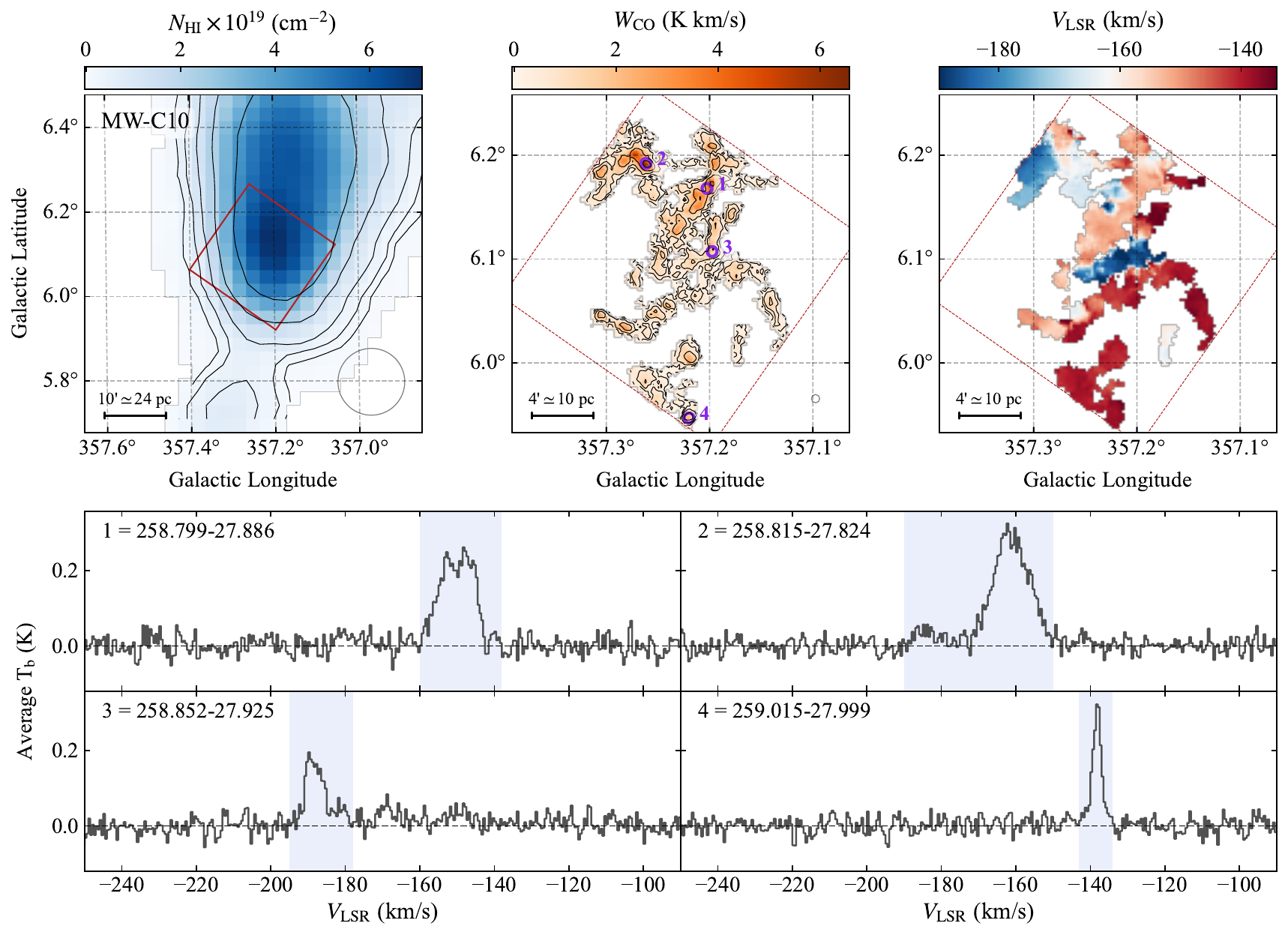}
    \vspace{0.5cm}
    \includegraphics[width=0.87\textwidth]{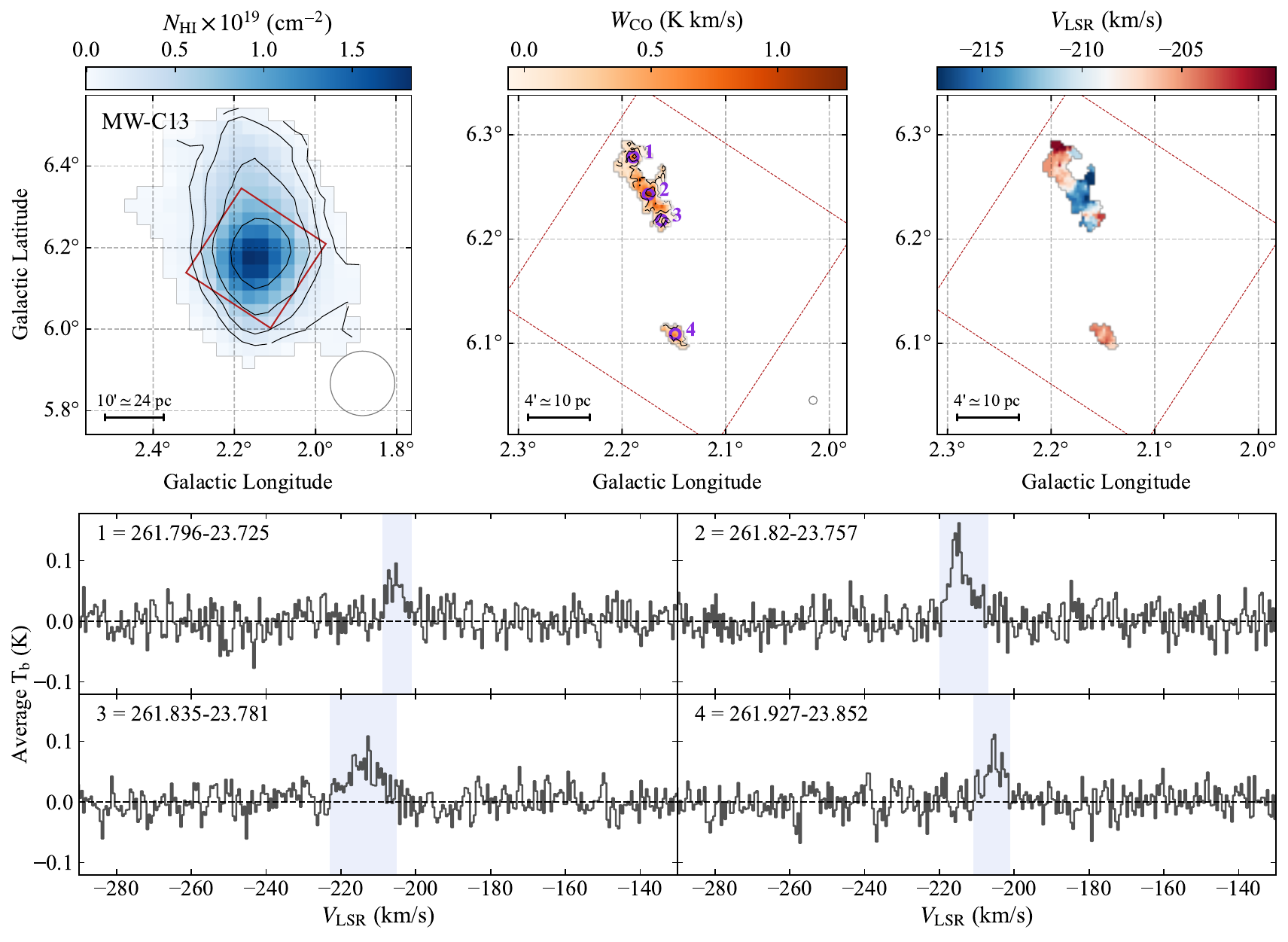}
    \vspace{-0.65cm}
    \caption{Moment maps and example spectra for two clouds (MW-C10 and MW-C13). For each cloud, we show the \hi\ column density map (\emph{left}, blue colorscale) from GBT data, the CO total intensity map (\emph{center}, orange colorscale) and the CO velocity map (\emph{right}, spectral colorscale). 
    The red squares denote the field observed in CO with APEX. 
    Beam sizes for the GBT and APEX data are indicated as grey circles in the bottom-right corner of the \hi\ and CO maps, respectively.
    Spectra are extracted from the positions indicated on the CO total maps, with radius equal to the APEX beam size. Blue shaded bands in the spectra denote the velocity range of the detection. Contour levels in the \hi\ and CO maps are at S/N levels of 2.5, 5, 10, 20 and 40.}
    \label{fig:twoclouds}
\end{figure*}

\begin{figure*}
    \centering
    \includegraphics[width=\textwidth]{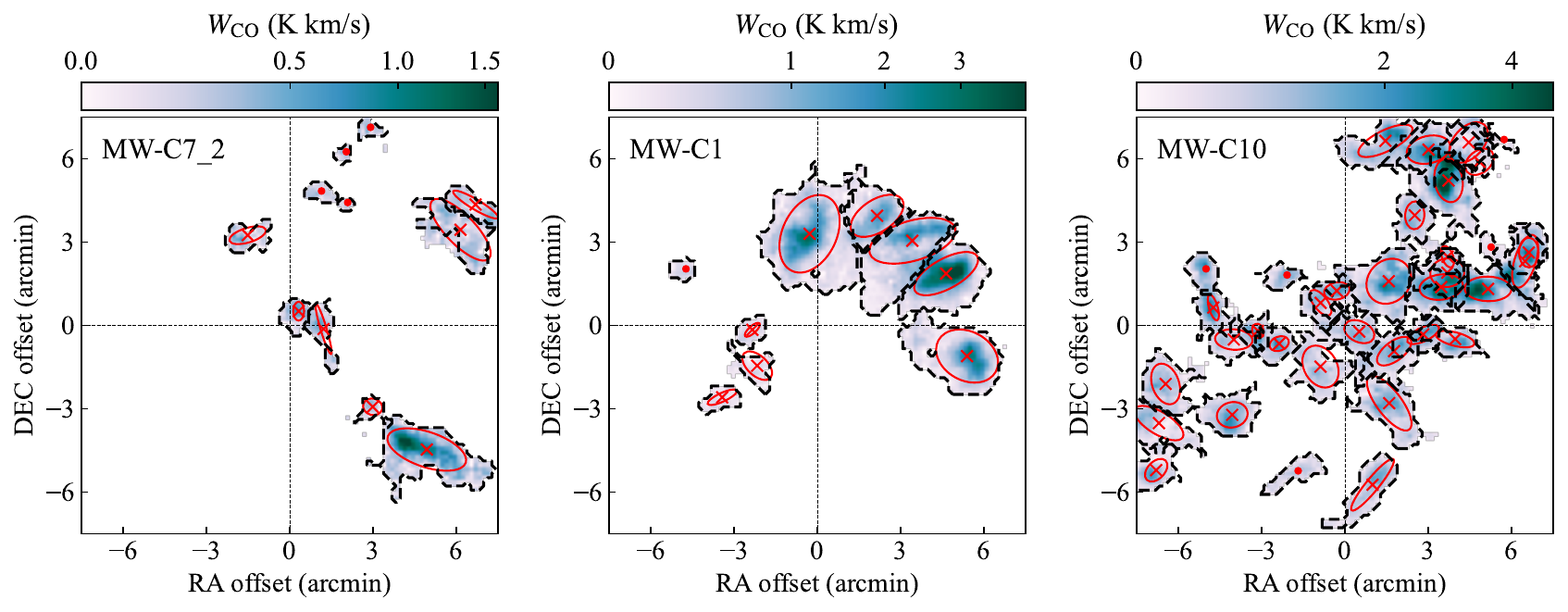}
    \caption{Cloud segmentation for three clouds in our sample: MW-C7\_2 (\emph{left}), MW-C1 (\emph{center}) and MW-C10 (\emph{right}). The blue-green colormaps show the masked \co\ integrated intensity maps.
    The black dashed contours denote the projected boundaries of each clump. 
    The red crosses and ellipses show the emission centroid and an elliptical approximation of the emission in each clump. 
    These ellipses represent a rough estimate of the size and orientation of the molecular clumps obtained through beam deconvolution. 
    Clumps that could not be deconvolved are indicated as a filled dot. 
    These clumps are not included in the analysis.}
    \label{fig:decomposition}
\end{figure*}

In summary, high-velocity \co\ emission is detected in 12 out of \cloudnum\ targets. 
Detected clouds are highlighted as full red circles in \autoref{fig:sample}, while non-detections are shown as dotted gray circles and indicated with a dagger in \autoref{tab:sample}.
MW-C12 and MW-C16 may have some very marginal signal at a $1.5\rms$ level, but being so weak we decided not to include them in the list of detections.
All CO sources have velocities consistent with the high-velocity \hi\ 
 cloud that was targeted.
In four fields (MW-C7, MW-C9, MW-C10, MW-C11), we also detect a secondary component at a different velocity with respect to the targeted \hi\ cloud: all these detections can be easily associated with additional high-velocity outflowing \hi\ clouds that coincidentally falls into the observed field of our CO data. 
These sources will be labeled with a $\_2$ throughout the paper (e.g. MW-C7\_2).
Taking also these secondary sources into account, we detect CO emission associated with a total of 16 \hi\ clouds in the GC outflow.

For each target, we create a 3D mask $\mathcal{M}(x,y,v)$ identifying the voxels belonging to a detection.
These masks are used during the analysis with \textsc{pycprops} and to derive moment maps for all detected CO sources through 3DB.
\autoref{fig:twoclouds} illustrates the APEX CO data for two clouds in our sample: MW-C10 (upper panels), a prototype of a cloud with abundant CO emission at high S/N, and MW-C13 (lower panels), a typical cloud with scarce and low S/N CO emission. 
For each cloud, we show a large-scale \hi\ column density map (top left) from the GBT data, the zero-th (integrated brightness temperature, top center) and the first (LSR velocity, top right) \co\ moment maps from APEX.
The lower contour level in the \hi\ and CO maps denote a S/N of 2.5, with the noise of the map calculated taking into account the masking applied to the cube \citep[see appendixes of][for details]{Verheijen+2001,Lelli+2014}.
Underneath the maps, we plot example spectra extracted from the four, beam-sized, circular regions indicated with numbered purple circles in the CO total map. 
These two clouds clearly show different characteristics from both a morphological and kinematical point of view. 
MW-C10 has strong CO emission, covering a large fraction of the observed APEX field, distributed along a filamentary-like structure. 
The molecular gas is spread over a quite large range of velocities, $-200 \, \kms \lesssim \vlrs \lesssim -120 \, \kms$, and the kinematics look very complex across the field.
On the contrary, MW-C13 shows only weak emission ($2-4$\rms) in two or three molecular gas clumps, restricted to a relatively small velocity range, $-220 \, \kms \lesssim \vlrs \lesssim -200 \, \kms$.
MW-C10 and MW-13 represent two extremes in our sample, with most of the other clouds having morpho-kinematics characteristics in between. 
Moment maps for all the clouds detected in this study can be found in \autoref{sec:appA}.

\subsection{Decomposition into CO clumps}
\label{sec:decomp}
Having determined the regions with CO emission in each field, the next step is to identify clumpy molecular gas cores and to derive their physical properties. 
To this end we make use of \textsc{pycprops}, an algorithm broadly-used for the segmentation and the analysis of resolved molecular clouds both in the Milky Way and in nearby galaxies.
\textsc{Pycprops} identifies distinct local maxima in a spectral cube and assigns them to different clouds/clumps. 
Maxima identification is performed through the \textsc{Astrodendro} package, which uses a dendogram representation \citep{Rosolowsky+2008} of the emission within the mask $\mathcal{M}(x,y,z)$.
These maxima are then used as a seed for a watershed algorithm that uniquely assigns remaining pixels within the mask (i.e., fainter emission) to a identified peak based on neighboring criteria \citep[see][for details]{Rosolowsky+2006,Rosolowsky+2021}.

For most of our data, we use a standard set of parameters for cloud decomposition. 
All pixel clusters assigned by the dendogram algorithm to local maxima were required to have a $2\rms$ contrast against the local noise level (\textsc{delta} parameter) and to cover a number of pixels at least equal to half size of beam (\textsc{minpix} = 6 pixels). 
A minimum separation of 45'' (\textsc{friends} = 5 pixels) and $2$ $\kms$ (\textsc{specfriends} = 4 channels) between different clumps was also required. 
Finally, we set a \textsc{compactness} parameter \citep{Neubert+2014} of 100, as a tradeoff between the need of identifying both compact and slightly extended molecular clumps.
A visual inspection of the resulting decompositions showed that these parameters are appropriate for the majority of our clouds. 
However, for the most complex and filamentary molecular structures (MW-C2, MW-C10 and MW-C15), we had to tweak the above parameters, slightly increasing the \textsc{delta} and \textsc{minpix} parameters.
This was necessary to avoid an excessive breaking down of the emission in these structures.

\autoref{fig:decomposition} shows the results of the clump decomposition for three clouds of increasing emission complexity: MW-C7\_2, MW-C1 and MW-C10. 
Overlayed to the CO integrated intensity maps, the black dashed contours show the projected boundaries of detected clumps for each \hi\ cloud. 
These projected boundaries can cross each other in the 2D map as the clumps have different velocities in the spectral cube. 
As shown in the figure, \textsc{pycprops} efficiently segments the emission in data where the emission is clearly clumpy and fairly regular, like in MW-C1. 
For the few clouds with a more complex morphology and/or kinematics, like the case of MW-C10, the algorithm still produces an acceptable decomposition. 
However, it is problematic to clearly define the boundaries between clumps in some regions where the emission morphology looks more similar to a filament than to a compact clump. 

Most of the 16 \hi\ clouds detected in CO have only a few ($5-10$) distinct molecular gas clumps each, like in the examples shown in the first two panels of \autoref{fig:decomposition}.
Instead, the three most complex clouds in our sample (MW-C2, MW-C10 and MW-C15) reach between 20 and 30 segmented molecular gas clumps.

\subsection{Physical properties of CO clumps}
Some important physical properties are calculated by \textsc{pycprops} for each molecular clump separately. 
These include the position and velocity centroid, linewidths, size, orientation and CO integrated intensity and luminosity.
Here we summarize how these quantities are estimated from the data \cite[for details, we refer to][]{Rosolowsky+2006}. 

The position and central velocity of a clump are simply its intensity-weighted mean along the two spatial dimensions (right ascension and declination) and along the spectral dimension, respectively. 
The velocity dispersion $\sigv$ (line width) is the square root of the intensity-weighted variance in the spectral direction (i.e. the second moment), taken over all the voxels that belong to a clump. 
The cloud size and orientation are calculated from the intensity-weighted second moments over the two spatial axes: the major and minor axes, $\sigma_\mathrm{maj}$ and $\sigma_\mathrm{min}$, and the position angle $\phi$ of the emission distribution are determined from the diagonalization of the spatial variance-covariance matrix. 
The measured cloud radius is then calculated as $R = \eta\sqrt{\sigma_\mathrm{maj}\sigma_\mathrm{min}}$, where we assumed the classic value of $\eta=1.91$ from \citet{Solomon+1987}.
Red ellipses in \autoref{fig:decomposition} show an example of the approximated elliptical representation of the segmented clumps, given by the measured $\sigma_\mathrm{maj}$, $\sigma_\mathrm{min}$ and $\phi$.

Finally, the CO luminosity of a given clump $\mathcal{C}$ is:

\begin{align}
    \label{eq:luminosity}
    \frac{L}{[\mathrm{K\,\kms \, pc^2}]} &= 
    \frac{D^2}{[\mathrm{pc^2}]} \sum_{i\,\in\, \mathcal{C}} \frac{T_i \, \Delta x \, \Delta y \, \Delta v}{[\mathrm{K\,\kms}]} = \\
    &= \frac{A_\mathrm{pix}}{[\mathrm{pc^2}]} \sum_{i\,\in\, \mathcal{C}} \frac{T_i\,\Delta v}{[\mathrm{K\,\kms}]} \nonumber
\end{align}

\noindent where $A_\mathrm{pix} \simeq (D^2\,\Delta x\,\Delta y)$ is the physical size of a pixel in pc$^2$, $\Delta x$ and $\Delta y$ are the pixel sizes in radians ($9''$ for CO APEX data and $115''$ for \hi\ GBT data) and $D=8200$ pc is the distance to the GC. 
$T$ is the brightness temperature in K and $\Delta v$ is the velocity width of a channel in $\kms$, and the sum is taken over all voxels in the clump. 

Uncertainties on the derived properties are calculated through a bootstrapping method, where several trial clouds are generated from the original cloud data and parameters are recomputed
for each trial cloud. 
In particular, for this work, we use 5000 realizations and calculate the parameter values and their uncertainties as the mean and standard deviation of the resulting parameter distributions. 
Finally, \textsc{pycprops} takes also into account biases introduced by the limited sensitivity and resolution of the observations. 
Some measured properties, like the spatial second moments and the luminosity, are extrapolated to correct for the missed flux due to the S/N clipping applied during cloud decomposition. 
Moreover, spatial and spectral resolutions have an effect on the estimates of the cloud size and line width, respectively. 
The rms beam size, $\sigma_\mathrm{beam}$, is subtracted in quadrature from the extrapolated spatial moments to obtain ``deconvolved'' quantities. 
In a similar way, the equivalent Gaussian width of a channel is subtracted in quadrature from the extrapolated velocity dispersion.

\subsection{Neutral and molecular gas masses}
\label{sec:masses}
Masses of neutral and molecular gas are estimated from the \hi\ and \co\ line emission, respectively. 
For 3D emission line data $T(x,y,v)$, the gas column density $N(x,y)$ at a sky position $(x,y)$ can be written as:

\begin{equation}
    \label{eq:column}
    \frac{N (x,y)}{\mathrm{[cm^{-2}}]} = \frac{X}{[\mathrm{cm^{-2} \, (K\, \kms)^{-1} }]} \sum_{v \, \in \, \mathcal{C}} \frac{T(x,y)(v)\,\Delta v}{[\mathrm{K \, \kms}]} 
\end{equation}

\noindent where the sum is taken over all velocity channels belonging to the clump $\mathcal{C}$ at the position $(x,y)$. 
The constant $X$ for the neutral gas emitting in \hi\ is $X_\hi = 1.82 \times 10^{18} \, \xcoun$, under the reasonable assumption that the gas is optically thin \citep[e.g.,][]{Roberts75}. 
For molecular gas emitting CO lines, the constant $X_\mathrm{CO}$ is known as the CO-to-H$_2$ conversion factor \citep[e.g.,][]{Bolatto+13}.
Our choice for the $X_\mathrm{CO}$ is based on radiative transfer modelling and is discussed extensively in Section~\ref{sec:radmod}.

The total ``luminous'' mass of gas will then be:

\begin{align}
    \label{eq:mass}
    \frac{M}{\mathrm{[M_\odot]}} &= 1.36 \, \frac{m}{\mathrm{[M_\odot]}} \, \frac{D^2}{\mathrm{[cm^2 \, ]}} \sum_{(x,y) \, \in \, \mathcal{C}} \frac {N(x,y)\, \Delta x \, \Delta y}{[\mathrm{cm^{-2}}]}  = \\
    &= \frac{\alpha}{\mathrm{[M_\odot \, (K \, \kms \, pc^2)^{-1}]}} \, \frac{L}{\mathrm{[K \, \kms \, pc^2]}} \nonumber
\end{align}

\noindent where the sum is made over the spatial pixels of a cloud, $m$ is either the mass of atomic/molecular hydrogen for atomic/molecular gas and the factor 1.36 is to take into account Helium. 
The last part of Eq.~(\ref{eq:mass}) is obtained by using Eqs.~(\ref{eq:luminosity}-\ref{eq:column}). 
The constant $\alpha = 1.36\,m\,X$ has the meaning of a mass-to-light ratio and converts the line luminosity $L$ to a mass. 

For all CO clumps isolated in each \hi\ cloud, we calculated the luminous masses from the CO luminosity. 
Unfortunately, the significantly coarser spatial resolution of our \hi\ GBT data ($570''$ vs $28''$) does not allow us to compare atomic and molecular gas contents for each single clump. 
High-resolution interferometric \hi\ observations currently exist only for two clouds with detected CO emission \citep[MW-C1 and MW-C2,][]{Noon+2023}.
Therefore, for each cloud, we calculate the \hi\ mass over the entire field covered by the APEX observations and we compare it to the total molecular gas mass in the same region.
For the 7 clouds with no CO detections, we estimate a molecular gas upper limit in the following way: we assume the existence of at least one molecular clump with a line peak at $3\rms$ and emission spread over a 3D Gaussian having spatial FWHMs equal to the beam size and the spectral dispersion set to 2 km/s (4 channels).
We highlight that this should be a quite conservative upper limit, given that such a clump would be detectable with our procedure, although barely. 

\begin{table}
\centering
\caption{Range of cloud and environmental parameters explored in the radiative transfer modelling with DESPOTIC.}
\label{tab:parspace}
\begin{tabular}{ll}
\noalign{\vspace{5pt}}\hline\hline\noalign{\vspace{5pt}}
Parameter\hspace{30pt} & Range or values \\
\noalign{\vspace{5pt}}\hline\hline\noalign{\vspace{5pt}}
$\log \, (n_\mathrm{H} / \mathrm{cm^{-3}})$ & $[1 - 4]$ with 50 steps of 0.06 dex\\
$\log \, (N_\mathrm{H} / \mathrm{cm^{-2}})$ & $[19-23]$ with 50 steps of 0.08 dex\\
$\sigma_\mathrm{turb}$ / $\kms$ & $[1, 3, 5, 7.5, 10]$ \\
$\log \; (\chi / \chi_0) $ & [-1, 0, 1, 2] \\
$\log \; (\zeta / \mathrm{s^{-1}})$ & [-17, -16, -15, -14] \\
$Z_\mathrm{dust} \; / \; Z_\mathrm{\odot} $ & [1, 2, 3] \\
\noalign{\vspace{3pt}}\hline \noalign{\vspace*{3pt}}
\end{tabular}
\end{table}

\section{Radiative transfer modelling and $X_\mathrm{CO}$}
\label{sec:radmod}
It is not trivial to convert the CO observed brightness temperature to a molecular mass for a cloud in a galactic wind environment. 
Unlike regular molecular clouds in galactic disks, the dynamical and chemical state of molecular gas in an outflow are unconstrained and we can not simply apply the usual Galactic conversion factor of 
$X_\mathrm{CO(1\rightarrow0)} = 2 \times 10^{20}$ $\xcoun$ \citep[e.g.,][]{Bolatto+13}.
Our early observations of three clouds in the wind \citep{DiTeodoro+2020,Heyer+2025} have also suggested a wide range of possible $X_\mathrm{CO}$ for these objects.

For this work, we use the radiative transfer software DESPOTIC \citep[Derive the Energetics and SPectra of Optically Thick Interstellar Clouds,][]{Krumholz14} to explore the range of physically-motivated conversion factors for clouds in a wind.
DESPOTIC calculates the thermal and chemical equilibrium state of an optically-thick cloud subject to a given interstellar radiation field $\chi$ and a given cosmic-ray ionization rate $\zeta$.
To compute the chemical equilibrium, we adopt the H$-$C$-$O chemical network by \citet{Gong+17}, varying the dust abundance $Z_\mathrm{dust}$ from solar to super-solar values. 
In the thermal balance, DESPOTIC takes into account photoelectric heating from dust grains, heating by cosmic rays, and turbulent heating, while cooling is due to common atomic and molecular transitions (including \hi, C\textsc{ii}, C\textsc{i}, O\textsc{i}, and CO) and energy transfer via collisions between gas and dust.
Level populations are computed using an escape probability approach, where the escape probabilities are derived based on DESPOTIC’s assumption of spherical cloud geometry. 

\begin{figure*}
    \centering
    \includegraphics[width=\textwidth]{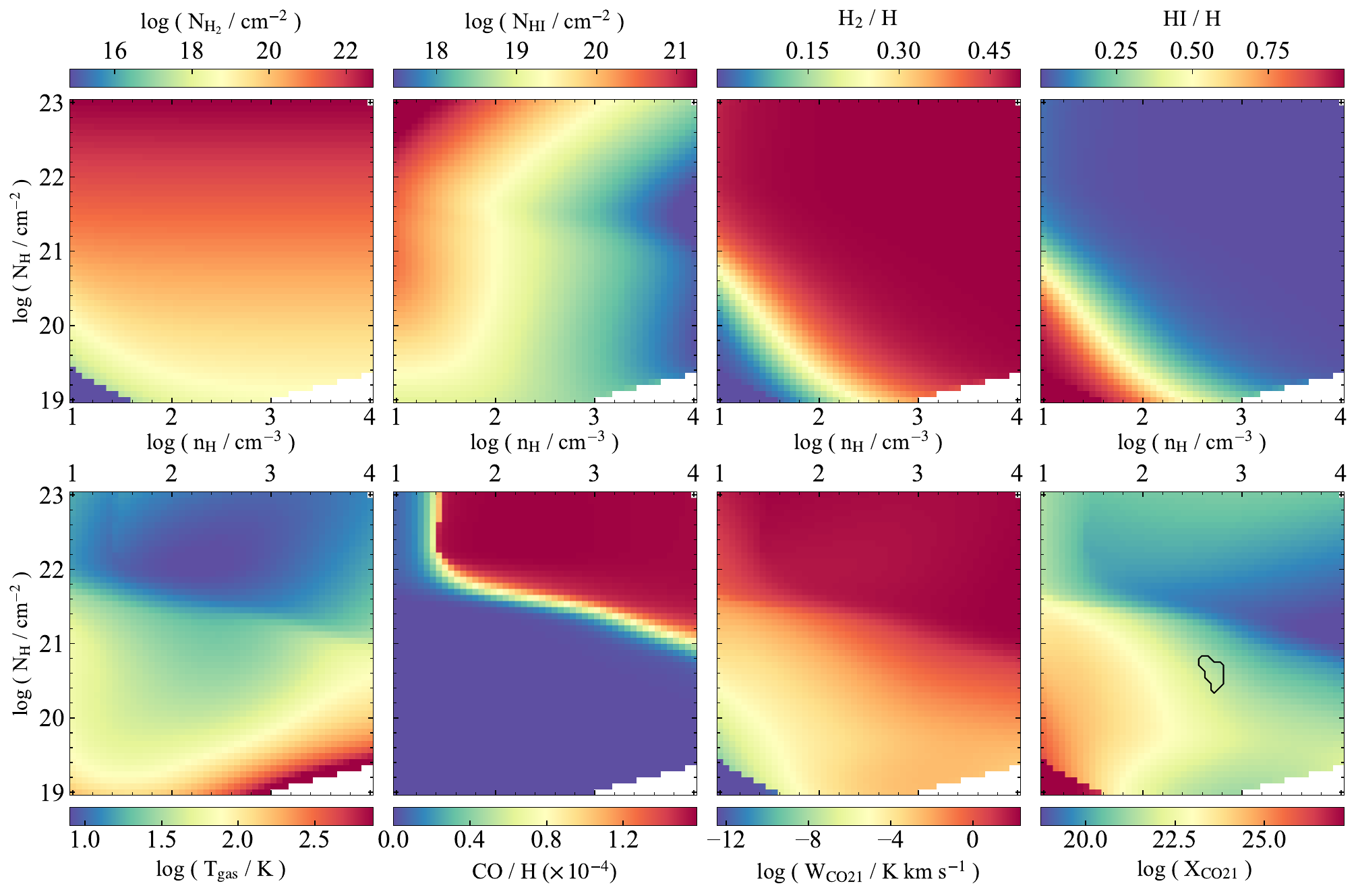}
    \caption{DESPOTIC results for a model with $\sigma_\mathrm{turb} = 3 \, \kms$, $\chi=\chi_0$, $\zeta=10^{-16}$ s$^{-1}$ and solar dust abundance. Top panels, from the left to the right: predicted H$_2$ column density, \hi\ column density, H$_2$ and $\hi$ abundances per H nucleus. 
    Bottom panels, from the left to the right: gas temperature, CO abundance per H nucleus, integrated brightness temperature and CO-to-H$_2$ conversion factor for the \co\ emission line. 
    Blank pixels denote regions of the grid where the models did not converge. 
    The black contours on the $X_\mathrm{CO}$ plot represent an example of the constraint on the parameter space given by our data for MW-C1. Similar radiative-transfer models are used to calculate fiducial conversion factors within each \hi\ cloud. 
    }
    \label{fig:models}
\end{figure*}

We approximate a molecular clump as a single-zone spherical cloud described by three parameters: its hydrogen number density $n_\mathrm{H}$, column density $N_\mathrm{H}$ and turbulent (i.e.\ non-thermal) velocity dispersion $\sigma_\mathrm{turb}$. 
For a given set of cloud parameters ($n_\mathrm{H}$, $N_\mathrm{H}$, $\sigma_\mathrm{turb}$) and environmental parameters ($\chi$, $\zeta$, $Z_\mathrm{dust}$), DESPOTIC computes the equilibrium state and returns the final thermodynamical/chemical quantities (e.g.\ gas temperature, H$_2$/\hi\ volume and column densities, abundances of the different species).
It also estimates the integrated brightness temperatures for the tracked species, e.g.\ $^{12}$CO and $^{13}$CO, and for different line transitions, e.g.\ $1\rightarrow0$ and $2\rightarrow1$. 
We perform these calculations on a $50\times50\times5\times4\times4\times3$ six-dimensional grid, exploring both the cloud and the environment parameter spaces.
In particular, we explore clouds with number densities $n_\mathrm{H} = 1 - 10,000$ cm$^{-3}$, column densities $N_\mathrm{H} = 10^{19} - 10^{23}$ cm$^{-2}$ and velocity dispersions $\sigma_\mathrm{turb} = 1 - 10$ $\kms$. 
We also explored different environments, varying the interstellar radiation field $\chi$ and the cosmic-ray ionization rate $\zeta$. 
The former is varied between 0.1$\chi_0$ and 1000$\chi_0$, where $\chi_0$ is the standard Solar radiation field \citep{Draine78}, while the latter between $10^{-17}$ s$^{-1}$ and $10^{-14}$ s$^{-1}$. 
We remind the reader that observations in the Solar neighborhood found that the ionization rate is $10^{-16}-10^{-17}$ s$^{-1}$ in diffuse and dense clouds, respectively, with quite a large dispersion around these values \citep[e.g.,][]{Padovani+2020}, while estimates in the CMZ span an even wider range, from $\sim 10^{-14}$ s$^{-1}$ \citep{Oka+19} to values not vastly different from those in the Solar neighborhood \citep{Ginsburg16a, Krumholz23a} and to values intermediate between these two \citep{Armillotta+20}. 
We consider dust abundances ($Z_\mathrm{dust}$) equal to one, two  and three times the Solar value.
With our choice of $\chi$, $\zeta$ and $Z_\mathrm{dust}$, we aim to probe a wide range of possible environments, from sub-Solar to 
highly super-Solar, to overcome our ignorance on the actual ambiance experienced by the clouds. 
On the one hand, because they are at GC and above the CMZ, we may expect a more extreme environment than the Solar neighborhood. 
On the other hand, our CO clouds do not lie in the Galactic disk, but they are at a few hundreds parsec to 2 kpc from it; 
because both the radiation field and the ionization rate are expected to drop with distance from the plane, models with intermediate $\chi$ and $\zeta$ might be more representative of the conditions in the wind.
The full parameter space explored with DESPOTIC is summarized in \autoref{tab:parspace}.

\autoref{fig:models} illustrates an example of outputs from DESPOTIC on the $\log(n_\mathrm{H})-\log(N_\mathrm{H})$ grid for a model with $\sigma_\mathrm{turb} = 3 \, \kms$, $\chi=\chi_0$, $\zeta=10^{-16}$ s$^{-1}$ and $Z_\mathrm{dust} = Z_\mathrm{\odot}$. 
The first row shows the predicted H$_2$ and \hi\ column densities and abundances: as expected, moving from low to high density clouds, the fraction of H$_2$ increases while that of \hi\ decreases. 
In this particular environment condition, a cloud is always fully molecular for $\log(n_\mathrm{H} / \mathrm{cm^{-3}}) \gtrsim 2.5$.
In the second row, we plot the gas temperature, the CO abundance, the integrated brightness temperature $W_\mathrm{CO(2-1)}$ for the \co\ line and the corresponding conversion factor $X_\mathrm{CO(2-1)} = N_\mathrm{H_2} / W_\mathrm{CO(2-1)}$, where the H$_2$ column density is calculated as $N_\mathrm{H_2} = x_\mathrm{H_2} N_\mathrm{H} $ with $x_\mathrm{H_2}$ being the H$_2$ abundance per hydrogen nucleus. 
From the last plot, we can notice that $X_\mathrm{CO(2-1)}$ spans over more five order of magnitude.
However, we can further constrain the $X_\mathrm{CO(2-1)}$ parameter space by using some observed properties of our data: in particular, the \hi\ column densities calculated from the \hi\ data through \autoref{eq:column}, the $W_\mathrm{CO}$ measured from the CO data and the cloud radius $R = 3/4 \, n_\mathrm{H_2} / N_\mathrm{H_2} $ obtained from the clump decomposition.
For example, the black contour on the $X_\mathrm{CO(2-1)}$ plot denotes the allowed parameter space after we impose the observed properties of clumps in MW-C1, i.e. $0.5 < W_\mathrm{CO} / ( \mathrm{K \, \kms}) < 4 $, $19 < \log (N_\mathrm{HI} / \mathrm{cm^{-2}}) < 21.5 $ and $R<4$ pc. 
Thus, for this particular model and observed cloud, the range of allowed $X_\mathrm{CO(2-1)}$ is reduced to $8\times10^{20}-5\times10^{21} \, \xcoun$.

For each cloud in our sample, we select the closest models in terms of $\sigma_\mathrm{turb}$ and we impose the observational clump measurements in $W_\mathrm{CO}$, $R$ and $N_\hi$. 
Adding up all the allowed values from all the models of the different environments, we build up distributions of $X_\mathrm{CO}$ for each cloud in the wind. 
From these $X_\mathrm{CO}$ distributions, which are fairly log-normal, we take the 50$^\mathrm{th}$ percentile as the fiducial value and the 15.87$^\mathrm{th}$ and 84.13$^\mathrm{th}$ percentiles as lower and upper uncertainty bounds.
This approach returns central values of $X_\mathrm{CO(2-1)}$ in the range $6 - 20 \times 10^{20} \, \xcoun$, several times larger than the typical Galactic conversion factor.
We use these values to calculate molecular masses for each CO clump, propagating the asymmetric errors with a Monte Carlo method.

Finally, we stress that our approach is model-dependent and relies on the assumption of thermal and chemical equilibrium, which may not apply to clouds entrained in galactic winds (see discussion in Section \ref{sec:mass}). 
\citet{Noon+2023} showed that at least two of our clouds (MW-C1 and MW-C2) have atomic shielding layers too thin to maintain chemical equilibrium. 
Non-equilibrium abundances can cause values of $X_\mathrm{CO}$ somewhat higher than those found in equilibrium, because when a cloud undergoes a rapid reduction in shielding (for example because its outer layers are stripped off as it is entrained into a wind), the resulting increase in UV flux affects the CO and H$_2$ on different timescales. The CO will rapidly find a new equilibrium when exposed to an increased radiation field, while self-shielding effects mean that it takes finite time for the increased photon flux to dissociate the H$_2$ and drive its abundance to a new, lower equilibrium. During this adjustment period, the CO luminosity is reduced, but the H$_2$ abundance has not yet caught up, increasing $X_\mathrm{CO}$, though the strength of this effect is still under investigation (Noon et al., in prep.). 

\begin{figure*}
    \centering
    \includegraphics[width=\textwidth]{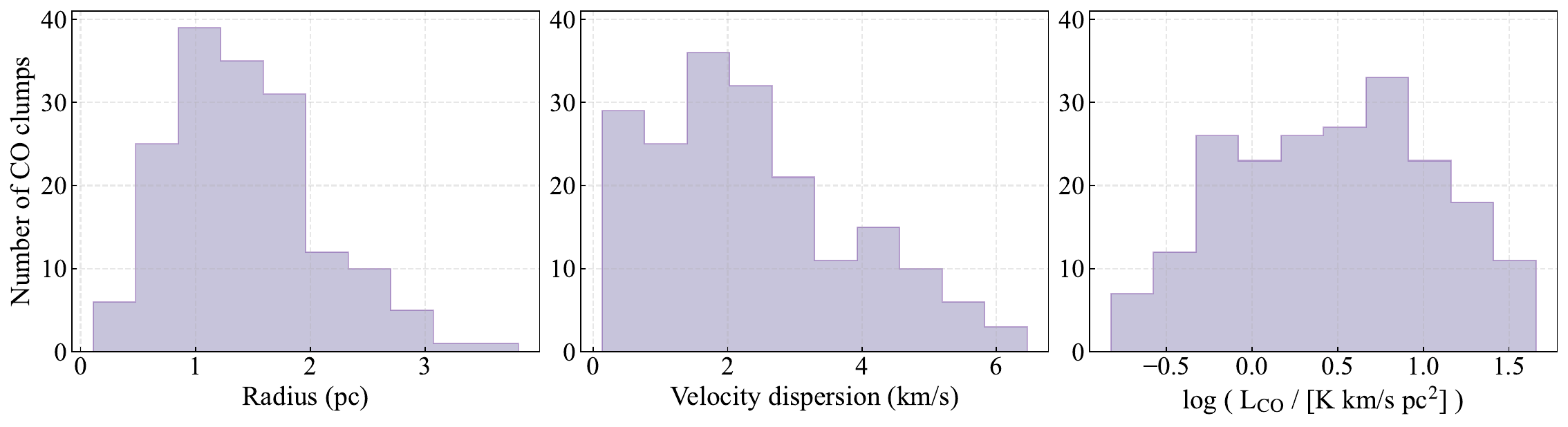}
    \caption{Histogram distributions of the observed properties of the molecular clumps detected in the APEX data. From the left to the right: cloud radius (assuming a distance of 8.2 kpc), velocity dispersion and \co\ luminosity (Equation~\ref{eq:luminosity}).}
    \label{fig:cloudprop}
\end{figure*}

\vspace{-0.15cm}

\section{Results}
\label{sec:results}

\subsection{Observed properties}
Our cloud decomposition method detected a total number of 207 molecular gas clumps within the 16 high-velocity \hi\ clouds with CO detections. As mentioned in Section~\ref{sec:decomp}, most \hi\ clouds have less than 10 segmented molecular clumps, but some of the most complex structures can show up to a few tens clumps (e.g., MW-C10). 
\autoref{fig:cloudprop} shows the distributions of some of the molecular clump properties: from the left to the right, we plot histograms of clump radii, line widths (velocity dispersion) and CO luminosities.
Detected clumps are compact, with typical sizes of $1-3$ pc, and a median value of $1.2$ pc.
They show velocity dispersions up to $\sigv\simeq6 \, \kms$, with a median value of 2 $\kms$, larger than the expected thermal broadening for a typical molecular cloud ($\lesssim 1 \,\kms$ for a temperature $T\lesssim 1000$ K), suggesting a significant level of turbulent broadening. 
Their CO luminosity ranges from 0.3 to 30 $\mathrm{K \, km/s \, pc^2}$, which is faint compared to regular clouds in the disk where CO luminosities range from 10 to $10^6$ $\mathrm{K \, km/s \, pc^2}$ \citep[][\citetalias{Miville-Deschenes+2017} hereinafter]{Miville-Deschenes+2017}.

\subsection{Scaling relations}
\begin{figure*}
    \centering
    \includegraphics[width=\textwidth]{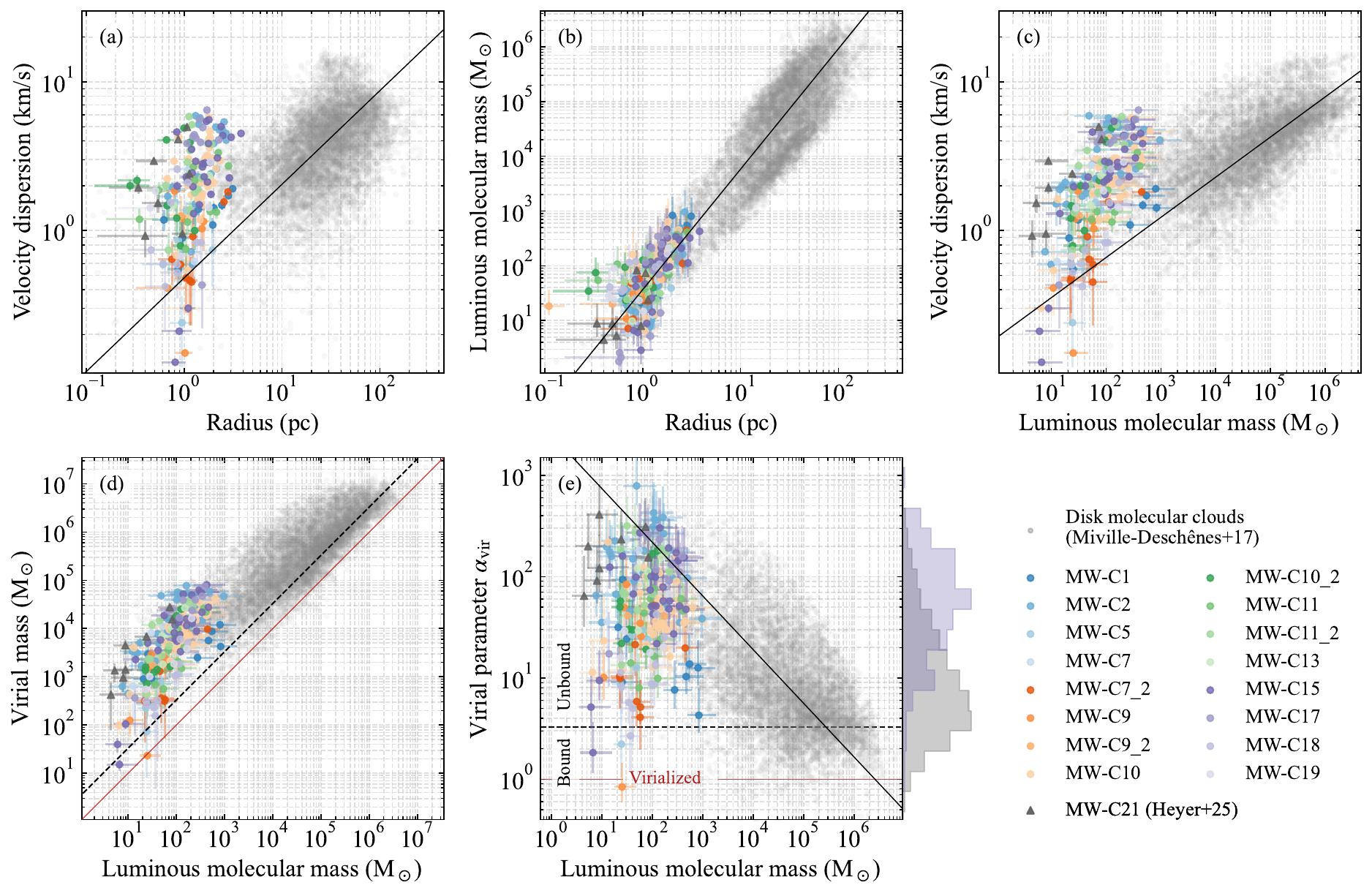}
    \caption{Scaling relations of molecular clouds in the MW wind compared to molecular clouds in the MW disk. 
    In all panels, grey dots indicate the MW disk sample from \citetalias{Miville-Deschenes+2017}, the points are the wind clumps studied in this paper from APEX data, color-coded by their parent \hi\ cloud, while the dark grey triangles are for the MW-C21 cloud from LMT data \citep{Heyer+2025}. 
    Full black lines denote the best-fit relations to the disk sample from \citetalias{Miville-Deschenes+2017}. Panel (a): $\sigv - R$ relation, best-fit $\sigv = 0.48R^{0.63}$; (b): $\mmol - R$ relation, best-fit $\mmol=36.7R^{2.2}$; (c) $\sigv - \mmol$ relation, best-fit $\sigv = 0.19\mmol^{0.27}$; (d) $\mvir$ vs $\mmol$; (e) $\avir$ vs $\mmol$, best-fit $\avir = 2500 \mmol^{-0.53}$. In panels (d) and (e), the full red line denotes a virialized cloud ($\mmol = \mvir$), while the black-dashed line indicates the $\avir = 3.3$ boundary between gravitationally bound and unbound clouds. 
    The histograms on the $y$-axis of panel (e) show normalized distributions of $\avir$ marginalized over $\mmol$ for disk clouds (gray) and wind clouds (purple). 
    The extremely large virial parameters, driven by the large velocity dispersions, imply that the clouds are not gravitationally bound.
    }
    \label{fig:scaling}
\end{figure*}

Molecular clouds in the MW and in nearby spiral galaxies are known to lie on scaling relations between their most important physical properties, namely radius, velocity dispersion, and mass \citep[e.g.,][]{Solomon+1987,Rosolowsky+2021,Deng+2025}.
We investigate whether the clumps defined in our study still obey the same relations.  
The top row of panels of \autoref{fig:scaling} illustrates these scaling laws for our cloud sample (colored points based on parent \hi\ cloud) compared to the MW disk measurements (grey dots) from \citetalias{Miville-Deschenes+2017}. 
In these plots, we also include the cloud MW-C21 detected in the $^{12}$CO(1$\rightarrow$0) line by \citet{Heyer+2025} (dark grey triangles), after reanalysing the LMT data with the same methods used for our APEX data for consistency.
From the left to the the right, we show the velocity dispersion - radius relation ($\sigv-R$),  the luminous mass - radius relation ($\mmol - R$) and the luminous mass - velocity dispersion relation ($\sigv - \mmol$). 
Full black lines denote the best-fit relations to the disk sample, indicated in the caption. 

Wind molecular clumps closely follow the $\mmol - R$ relation (panel b) of disk molecular clouds, extending the power law to the low-mass small-radius regime. 
Radius and mass are clearly correlated quantities for our sample, with a Spearman rank-order correlation coefficient of $\rho\sim0.7$ (with 1 indicating perfect monotonically-increasing correlation and 0 no correlation).
We note that the choice of conversion factors, based on DESPOTIC modeling, is key to bringing our clouds onto the relation. Had we used a standard Galactic $X_\mathrm{CO}$, the wind clouds would have lain about an order of magnitude below the relation. 
The fact that wind clouds lie on the mass - size relation ($\mmol \propto R^2$) indicates that these objects have nearly constant average surface densities that do not depend on $R$, which also implies that $n_{H_2} \propto R^{-1}$ \citep[e.g.,][]{Larson+1981}. 
This is also a general, well-known feature of molecular clouds in the MW disk \citep[e.g.,][]{Roman-Duval+2010,Chen+2020,Lewis+2022}.
The intercept of the relation implies an average surface density of $\sim 40$ $\mo$ pc$^{-2}$, which for Solar dust abundances corresponds to a visual extinction $A_V$ of a few magnitudes, roughly the extinction theoretically predicted to be necessary to shield CO molecules from photodissociation by the interstellar radiation field \citep{Bolatto+13}. Thus the fact that the wind and Galactic plane molecular clumps both feature constant surface densities of this value likely reflects that for both classes of object the edge of the CO-bright region is determined by the common physics of photodissociation.

Unlike the $\mmol - R$ relation, the clumps do not follow neither the $\sigv - R$ (panel a in \autoref{fig:scaling}) nor the $\sigv - \mmol$ (panel c) relation of disk clouds: in both cases, wind clouds have larger velocity dispersions than regular molecular clouds, and lie well above the relations, in a region populated by only a handful disk clouds. 
The $\sigv - R$ relation is the only one based on pure observational measurements (modulo an assumed distance), and does not depend on our choice of $X_\mathrm{CO}$, which is model dependent and remains the greatest source of uncertainty.
The $\sigv - R$ plane for disk clouds is known to show a large scatter and a relatively low correlation coefficient \citep[see e.g., the reviews by][]{Hennebelle&Falgarone2012,Heyer+15}. 
For our wind sample of clumps, we also find a large scatter and a Spearman coefficient $\rho\sim0.2$, indicating a very marginal positive correlation between radius and velocity dispersion, compared to a $\rho\sim0.5$ for the disk sample. 
Similarly, the $\sigv-\mmol$ plane for wind clouds also shows a similar weak correlation ($\rho\sim0.3$ compared to $\rho\sim0.7$ for MW disk clouds).

\subsection{Velocity dispersion and turbulence}
\label{sec:veldisp}

The large velocity dispersion of our sample suggests a higher level of turbulence in clouds in the wind with respect to clouds in the disk. 
A useful parameter in this context is the velocity dispersion normalized at 1 pc, $\sigma_0 = \sigv / (R / [1 \, \mathrm{pc}])^\gamma $, which basically represents the typical amplitude of turbulent motions on a parsec scale. 
The exponent $\gamma$ is related to the nature of the turbulence, $\gamma = 1/3$ for subsonic turbulence and $\gamma = 1/2$ for supersonic turbulence.
Assuming the latter, we find a mean $\sigma_0\simeq2 \, \kms$ for our sample, a number that is at least three times larger than in typical molecular clouds in the disk \citep[$\sigma_0\sim 0.4-0.8 \, \kms$, e.g.,][\citetalias{Miville-Deschenes+2017}]{Solomon+1987}.

Finally, we can estimate the internal (turbulent) pressure experienced by the clumps as

\begin{align}
   \label{eq:pressure}
   \frac {P_\mathrm{int}}{k} = \frac{\mu m_\mathrm{H}}{k} n_\mathrm{H_2}\sigv^2 = \frac{3 \mmol \sigv^2 }{4\pi k R^3}
\end{align}

\noindent where $k$ is the Boltzmann constant, $m_\mathrm{H}$ the hydrogen mass and $\mu$ the molecular weight.
Our wind clumps have turbulent pressures $P_\mathrm{int}/k \sim 10^4 - 10^7$ K cm$^3$. 
Molecular clouds in the MW disk have also a wide range of pressures, mostly between $10^3 - 10^6$ K cm$^3$, with the highest $P_\mathrm{int}$ found in the central kpc of the Galaxy and a decreasing trend as a function of Galactocentric radius \citepalias[][]{Miville-Deschenes+2017}.

\begin{figure*}
    \centering
    \includegraphics[width=\textwidth]{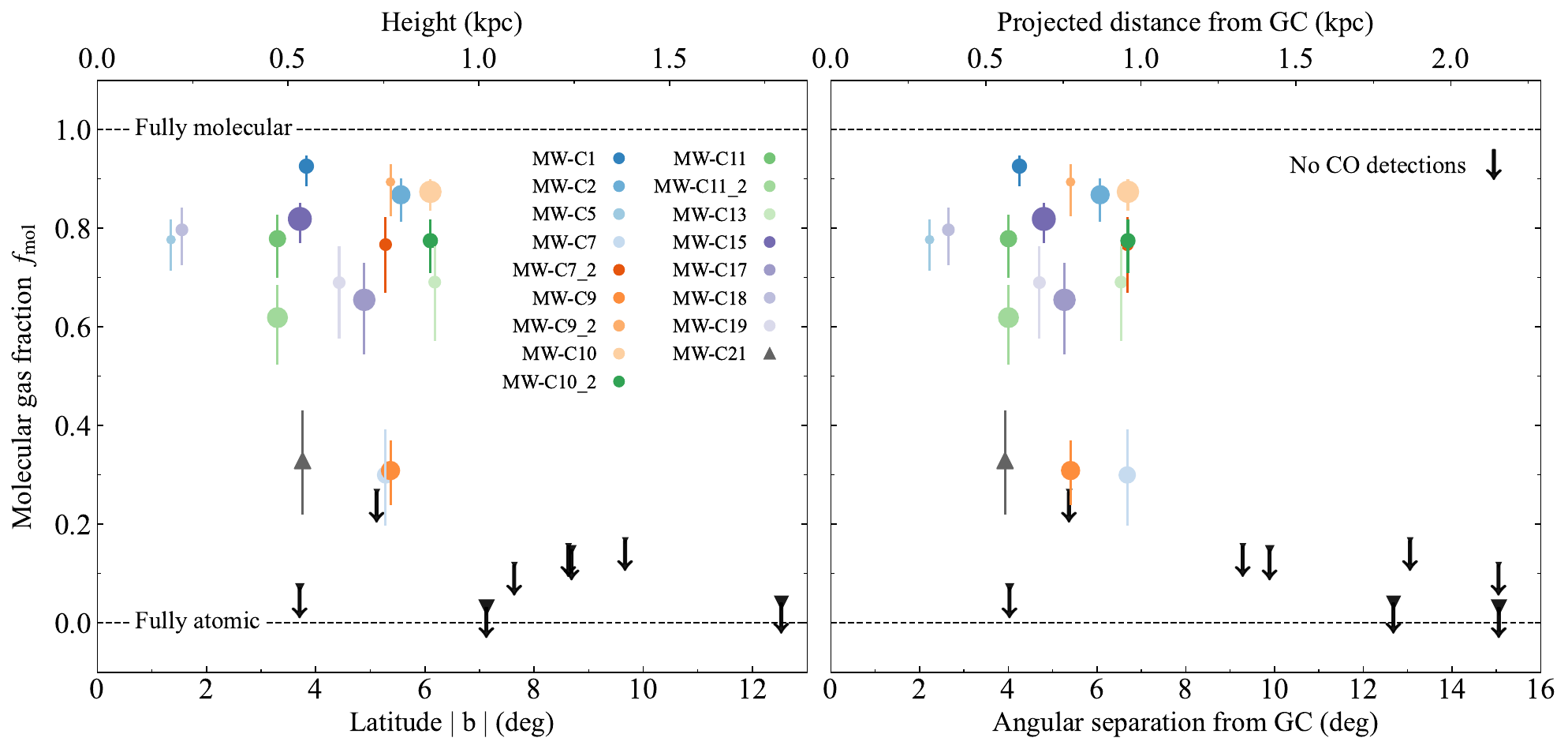}
    \caption{
    Molecular gas fraction as a function of latitude or height (left panel) and as a function of angular separation or projected distance from the Galactic Center (right panel). 
    Points denote the CO detections, color-coded as in \autoref{fig:scaling}. Up-down triangles with arrows represents observed \hi\ clouds with no CO detections, for which we estimate molecular mass upper limits (see Section~\ref{sec:masses}).
    The size of all markers is proportional to the peak \hi\ column density of the cloud, ranging from $5 \times 10^{18}$ cm$^{-2}$ to $8 \times 10^{19}$ cm$^{-2}$. 
    }
    \label{fig:fracevo}
\end{figure*}

\subsection{Virial ratio and clump boundedness}
\label{sec:virialpar}

Given the high internal pressures, one may wonder whether these clouds in the wind are gravitationally bound or not.
A broadly-used approach to investigate this makes use of the ``virial'' mass, i.e. the mass that a self-gravitating molecular cloud should have if it was in virial equilibrium. 
For a cloud with a spherical volume density profile $\rho(r) = r^{-\beta}$ and no magnetic support nor pressure confinement, the virial mass can be written as \citep{Solomon+1987}:

\begin{align}
    \label{eq:virmass}
    \mvir = \frac{3(5-2\beta)}{(3-\beta)} \, \frac{R \, \sigv^2}{G}
\end{align}

\noindent where $G$ is gravitational constant. 
In this work, we assume $\beta = 1 $ based on the findings of the $\mmol - R$ relation and for consistency with previous studies, which leads to the well known $M_\mathrm{vir,\, [\mo]} \simeq 1040 \, R_\mathrm{[pc]} \, \sigma^2_\mathrm{v,\,[km/s]}$. 
We note however that Eq.~(\ref{eq:virmass}) depends only weakly on the chosen value of $\beta$.
Deviations from virialization can be studied through the so-called virial parameter $\avir$, which roughly corresponds to the ratio of total kinetic energy to gravitational energy \citep{Bertoldi+1992}:

\begin{align}
    \label{eq:virpar}
    \avir = \frac{3(5-2\beta)}{(3-\beta)} \frac{R\, \sigv^2}{GM} = \frac{\mvir}{\mmol}\quad .
\end{align}

\noindent Gravitationally bound objects have $\avir\sim1$; 
large ratios, i.e.\ $\avir\gg1$, indicates that there is an excess of kinetic energy and thus a cloud would expand and possibly dissipate without further confinement; $\avir\ll1$ suggests that a cloud would collapse without further support.

In the bottom panels of \autoref{fig:scaling}, we show plots of luminous molecular mass vs virial mass (left) and of virial parameter vs luminous mass (right) for our cloud sample.
Full red lines denote $\mmol = \mvir$, i.e.\ the locus of virialized clouds. 
Most of our clouds have virial masses that exceed their luminous molecular mass by more than an order of magnitude, with corresponding virial parameters ranging from a few to several hundreds. 
This is mainly driven by the higher than usual velocity dispersions, as discussed in the previous Section.
Nonetheless, they seem to mildly follow the general trend, seen also in disk clouds, of a larger $\avir$ for low-mass clouds. 

In these plots, black-dashed lines represent $\avir = 3.3$, which approximately indicates the equilibrium between kinetic and gravitational energy \citepalias[see Appendix D in][for details]{Miville-Deschenes+2017}: clouds above (below) this line have more (less) kinetic than gravitational energy and can therefore be considered gravitationally unbound (bound).
While most of the molecular mass in the MW disk is confined within clouds that are either bound or loosely unbound ($\avir\lesssim10$), the vast majority (nearly 90\%) of wind clouds appear strongly unbound ($\avir\gg10-100$).
This can be appreciated by looking at the $\avir$ distributions for wind clouds (purple histograms) and disk clouds (gray), plotted in panel (e) of \autoref{fig:scaling}.
These characteristics make the wind clouds more similar to giant molecular clouds observed in galaxy centers \citep{Rosolowsky+2021} and high-velocity molecular clouds \citep{Nagata+2025} of external galaxies.

\subsection{Molecular mass fraction and evolution with latitude}
\label{sec:molfrac}

For each \hi\ cloud with a CO detection, we calculate the molecular gas mass fraction $\fmol$:

\begin{align}
   \label{eq:fraction}
   \fmol = \frac{\sum_i M^i_\mathrm{mol}}{\sum_i M^i_\mathrm{mol} + M_\mathrm{at}}
\end{align}

\noindent where $M_\mathrm{at}$ is the atomic gas mass derived from the GBT \hi\ data in a region coincident with the APEX field (thus not in the entire \hi\ cloud), and the sum for the molecular mass is taken over all the detected clumps in a \hi\ cloud. 
We calculate the fraction globally for each \hi\ cloud because, as mentioned in Section~\ref{sec:masses}, the mismatch between the spatial resolutions of GBT and APEX does not allow us to calculate it clump by clump.
For \hi\ clouds with no CO detections, we calculate an upper limit on the detectable molecular gas mass (see Section~\ref{sec:masses}), which translates into an upper limit on $\fmol$.
Most of our clouds with CO emission have very high molecular gas fraction, $\fmol=0.6-0.9$, meaning that 60\% to 90\% of the total gas mass is in the molecular phase. 
Only two clouds (MW-C7 and MW-C9) show lower gas fractions of $\fmol=0.3$.
The higher molecular fractions reported here, compared to those found in the previous works by \citet{DiTeodoro+2020} and \citet{Heyer+2025}, are due to the larger $X_\mathrm{CO}$ values adopted in this work.
A table with our derived masses and molecular gas fraction can be found in \autoref{sec:appB}.

In a scenario where cold gas is lifted up from the disk and entrained within the hot wind, we might expect the properties ($R$, $\sigv$ and $\fmol$) of the cold clouds to evolve with time and distance from the Galactic center, due to the combination of dynamical interaction between the hot and the cooler phases and photodissociation \citep{Vijayan24b}.
However, in our sample we see no clear signature of an evolution with distance of the observed clump parameters.
In particular, \autoref{fig:fracevo} shows the evolution of the molecular gas fraction as a function of latitude/height (left panel) and angular separation/projected distance from the GC (right panel). 
Markers and colors follow the convention of \autoref{fig:scaling}, while the up-down arrows denote the upper limits corresponding to non-detections in CO.
While there is no visible trend of $\fmol$ with height/distance, it is interesting to note that we have no CO detections in clouds above a latitude $| b | \gtrsim 7^\circ$, corresponding to a distance of about 1 kpc. 
We note that, although the APEX subsample was selected to broadly reflect the spatial and column-density distribution of the underlying \hi\ population, these latitude trends may still be influenced by completeness biases and should therefore be interpreted with caution.
Finally, we test whether the sharp decline in molecular gas could be linked to the existence of ionization cones proposed by \citet{Bland-Hawthorn+2019} \citep[see also][]{Sarkar2024}. 
In this scenario, a recent ($\sim3$ Myr) Seyfert-like burst from Sgr A$^*$ produced radiative cones that are tilted by $\sim 15\de$ with respect to South Galactic pole, potentially dissociating or ionizing gas within their volume. 
However, when examining cloud positions and molecular gas fractions as a function of angular offset from the cone axis, we find no correlation and no tendency for CO non-detections to lie near the axis.

One might expect that higher molecular gas fractions are preferentially found in clouds with higher \hi\ column density, as a dense atomic layer should provide shielding to molecule dissociation. 
Yet, in our sample we find no correlation between the $\fmol$ and the \hi\ column density.
This can be appreciated in \autoref{fig:fracevo}, where the size of the markers is proportional to the peak \hi\ column density: some clouds with very high \hi\ column densities (e.g., MW-C3 or MW-C16) have no detectable CO emission; even more puzzling, some clouds that are very faint in \hi\ show CO detections (e.g., MW-C5, MW-C14, see also \autoref{fig:sample}). 
If molecular clouds are entrained in the hot wind and gradually dissociate into atomic hydrogen \citep{Noon+2023}, 
these cases may represent opposite stages of the evolutionary sequence.
Clouds that have just become entrained may quickly lose their diffuse outer layers, stripping away much of the \hi\ while leaving dense molecular cores.
As these cores are subsequently exposed to dissociating radiation, they are destroyed from the outside in, producing clouds rich in \hi\ but poor in molecular gas.
However, we also stress that the large beam of the GBT ($570''$) smears out \hi\ emission and leads to measure lower column densities than the real ones; this beam dilution effect likely hides compact and high-density \hi\ clumps hosting molecular gas. 
High-resolution interferometric \hi\ data, matching the resolution of CO data, are essential to reliably study the relationship between neutral and molecular gas.
  
\section{Discussion}
\label{sec:discussion}

\subsection{What drives turbulence?}

Our results suggest that cold molecular clouds in the MW nuclear wind have physical properties that differ from regular molecular clouds in the disk.
Although they have masses and sizes generally smaller than the bulk of molecular clouds in the disk, they still lie on the mass - size relation. 
The most clear distinctive feature of wind clouds is their broad linewidths, generally larger than their disk counterparts at fixed size or mass, placing them well above the canonical linewidth - size relation and indicating an excess of non-thermal motions.
This implies a high level of turbulence and a large internal turbulent pressure, of the order of $P/k_{\rm B} \sim 10^4-10^7$~K~cm$^{-3}$.

One key open question is what drives the turbulence observed in these clouds. 
In the case of molecular clouds in the disk, several mechanisms have been proposed, such as stellar feedback, gravitational instabilities, or magnetorotational instabilities \citep{Heyer+15}, but these are unlikely to operate in the same way in clouds embedded in a galactic wind.
The clouds in our sample are typically located at heights of $0.2-1$~kpc above the Galactic plane, and are therefore relatively far from active star-forming regions in the disk and their associated feedback.
In addition, the absence of high-density gas tracers like HCN and HCO$^+$ in our deep APEX pointings toward some of the densest CO clumps supports the idea that these objects are diffuse and not sites of star formation.
Stellar feedback is therefore an unlikely source of turbulence, suggesting that other mechanisms must be responsible.
The interaction between the hot and cold phases of the outflow could be the main driver of turbulence, through compression and hydrodynamical instabilities at the interface between the two fluids. 
This picture is also supported by the recent findings of \citet{Gerrard+2024}: their analysis of the turbulence properties of MW-C1 and MW-C2 found that both clouds are in a sub-to-trans-sonic regime (Mach numbers $\mathcal{M}\simeq 1.1-1.4$) and exhibit predominately compressively-driven turbulence \citep[with a turbulence parameter $b\simeq1$;][]{Federrath+2010}, especially in the cloud's head.

\subsection{Cloud confinement}

Given the high internal pressure, one might expect molecular clouds embedded in the wind to expand and disperse over just a few clump crossing times.   
The crossing time for a cloud is approximately $t_\mathrm{exp} \sim R / \sigv$. 
For a typical cloud of $R\simeq1$ pc and $\sigv = 2 \, \kms$, this yields an expansion timescale of the order of $\sim0.5$ Myr, significantly shorter than the kinematic timescale of entrainment, estimated to be $\simeq 5-10$ Myr \citep{DiT+18,Lockman+20}.
For the clouds to survive over such timescales, some form of binding force must at least partially counteract their internal pressure. 
Our virial analysis shows that self-gravity alone is insufficient to bind the clouds, as we derive virial ratios $\avir\gg10-100$ for most objects in our sample. 

Clumps are embedded within \hi\ clouds, and both are entrained within the hot flow. 
External pressure from both these phases may help the molecular clumps to remain bound. 
Neutral gas can contribute with both thermal pressure, $P_\mathrm{th} / k = n T$, and turbulent pressure, given by Equation~(\ref{eq:pressure}).
For typical properties of the observed \hi\ population, with temperatures in the range $T\simeq1000-8000$ K, densities $n=0.1-10$ cm$^{-3}$ and velocity dispersions $\sigma_\mathrm{v,\, neu} \sim 5-10$ km/s, we expect thermal pressures of a few $10^4$ K cm$^{-3}$ and turbulent pressures of $10^4-10^5$ K cm$^{-3}$.
The physical conditions of the hot wind are more uncertain. 
Assuming number densities $n=10^{-3}-10^{-1}$ cm$^{-3}$ and temperatures $T=10^6-10^7$ K, the corresponding thermal pressures span $10^4-10^6$ K cm$^{-3}$.
In the most optimistic scenario, the total external pressure could therefore reach $10^6$ K cm$^{-3}$, which would be sufficient to confine approximately half of the observed molecular clumps in the wind. 
However, achieving such high external pressures would require an extremely dense ($\sim0.1$ cm$^{-3}$) and hot medium ($\sim10^7$ K) up to distances of $\simeq1$ kpc. 
These values are unlikely at these distances, as they exceed both theoretical predictions for hot winds \citep[e.g.,][]{Chevalier+1985,Bustard+2016} and observational estimates from X-ray studies in the MW \citep{Ponti+2019,Yeung+2024}.
A large fraction of these clouds may therefore be undergoing some degree of expansion, although in our data we do not see any strong trend in cloud radius as a function of distance from the Galactic plane. 
Additional confinement mechanisms, such as magnetic fields and cosmic ray pressure, may also be in place. 
These findings, based on a much larger sample of molecular clumps, fully confirm and strengthen the results of \citet{Heyer+2025}, who reached similar conclusions for the case of MW-C21.

\subsection{Molecular gas dissociation}

The absence of CO detections in the six \hi\ clouds located at distances larger than 1 kpc is also intriguing. 
While this may be a coincidence, if confirmed in future observations, it would suggest that molecular gas does not survive for long within the wind and is dissociated during its journey, as also proposed by \citet{Noon+2023}. 
For a cloud traveling at a constant velocity of $\sim 300 \, \kms$, it would take approximately 3 Myr to reach a distance of 1 kpc, providing a rough upper limit for the molecular gas dissociation timescale. 
However, since most clouds at heights of $0.5-1$ kpc still show very high molecular fractions ($\fmol\sim0.7-0.8$), the dissociation process may occur on much shorter timescales, possibly less than 1~Myr.

One speculative explanation for the presence of the sharp transition around $1$ kpc could be a change in the properties of the hot ambient medium surrounding the clouds: for instance, if the hot gas becomes significantly less dense and cooler above 1 kpc, the resulting drop in external pressure could trigger rapid cloud expansion.
This expansion might enhance the cloud exposure to the radiation field and to hydrodynamical instabilities, thereby accelerating the dissociation of molecular gas.
Clearly, further high-resolution observations of the neutral and molecular gas content in the MW wind are essential to develop a more comprehensive understanding of these processes.

\subsection{Mass loading and molecular mass uncertainties}
\label{sec:mass}

All the clouds with CO detections have high molecular gas fraction, ranging from 0.3 to 0.9, with a median value of 0.66.
The total atomic gas mass in the entire known \hi\ population at heights $z\leq 1$ kpc is $M_\mathrm{at,\,tot} \simeq 1.5\times10^5 \, \mo$.
If we assume that each of these clouds has a median molecular fraction of 0.66, the total cold (molecular plus atomic) gas mass within 1 kpc adds up to $M_\mathrm{cold,\,tot} \simeq 3\times10^5 \, \mo$.
For a timescale $t\simeq3$ Myr to reach 1 kpc at a velocity of 300 $\kms$, we can estimate a cold mass outflow rate of $\dot{M}_\mathrm{cold} = M_\mathrm{cold,\,tot} / \, t \simeq 0.1 \, \moyr$.
Given the star-formation rate of $\mathrm{SFR}\sim0.1 \, \mo$ in the CMZ \citep{Henshaw+2023}, we obtain cold mass loading factors $\eta = \dot{M}_\mathrm{cold} / \mathrm{SFR}$ of the order of unity at $0< z \, /\, \mathrm{kpc} < 1$.
Such a cold mass-loading factor is broadly consistent with the molecular loading factors reported for nearby starbursts ($\eta\sim0.5-5$) and lies at the lower end of the values measured in AGN-driven winds ($\eta\sim1-10$) \citep[see e.g.,][]{Fluetsch+2019,Veilleux+20}. We emphasize, however, that direct  robust comparisons remain difficult, as loading factor estimates are sensitive to differences in spatial resolution, adopted assumptions (e.g., conversion factor, outflow geometry and timescale), and the physical scales over which these measurements are made.

The dominant source of uncertainty in this work lies in the conversion between the observed integrated CO flux and molecular gas mass.
We attempt to constrain a range of plausible conversion factors using physically-motivated models through DESPOTIC (Section~\ref{sec:radmod}), although these rely on assumptions that may not be fully applicable to clouds embedded in a wind (i.e., thermal and chemical equilibrium).
Our modeling returns a median $X_\mathrm{CO21} \simeq 1 \times10^{21}$ $\xcoun$ across all the \hi\ clouds, corresponding to a median $\alpha_\mathrm{CO21} \simeq 22$  $\acoun$, which is about three times higher than the canonical Galactic value of 6.7 $\acoun$ for the \co\ transition \citep{Leroy+2009,Bolatto+13}.
While this Galactic $\alpha_\mathrm{CO21}$ is allowed within our radiative transfer models, it typically lies in the low-value tail of the resulting distributions, making it less likely. 
We emphasize that adopting a fixed Galactic conversion factor would not significantly alter our main conclusions.
Molecular gas masses and turbulent pressure would decrease by a factor of three, virial ratios would increase by the same factor. Molecular gas fractions would span the range $0.2-0.7$, with a median value around 0.35.
Our estimates of the total cold gas mass, mass outflow rate and loading factor would be reduced by approximately a factor of two.
Therefore, while absolute masses remain sensitive to the choice of $X_{\rm CO}$, our detection statistics, kinematics, virial and pressure analyses, and mass-loading estimates are robust to plausible conversion factor choices.
In any case, we would consistently find an observed population of molecular clumps that are gravitationally unbound and exhibit high levels of internal turbulent pressure.

\begin{figure}
    \centering
    \includegraphics[width=0.35\textwidth]{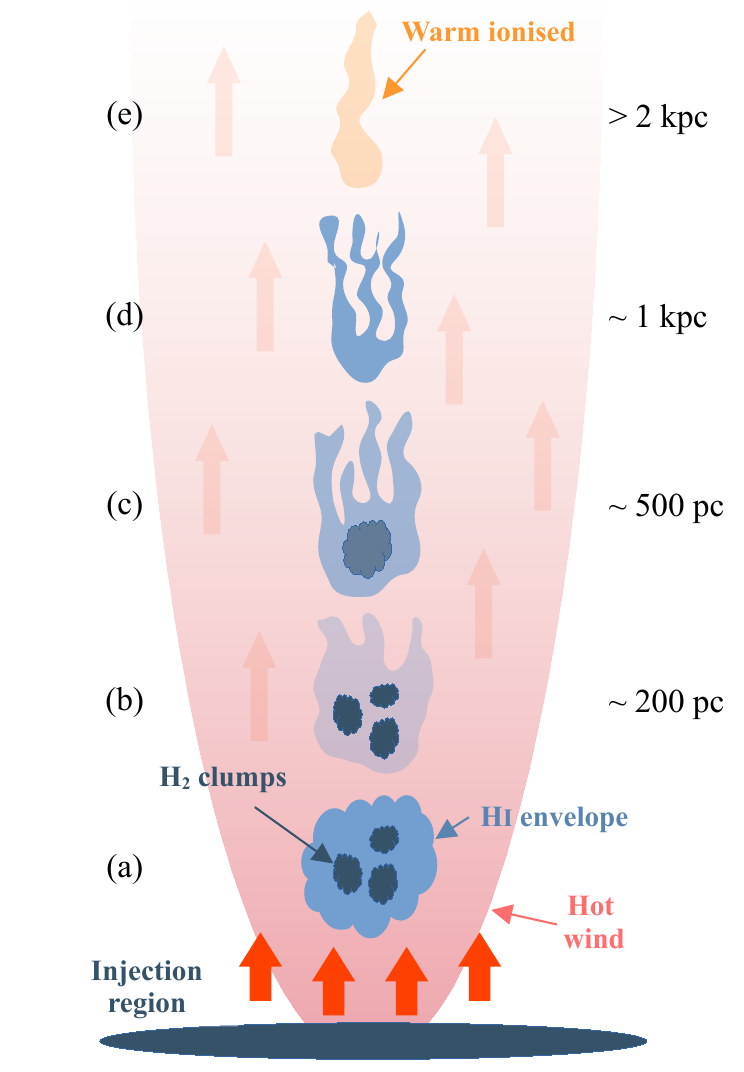}
    \caption{
    A cartoon illustrating a possible scenario for the evolution of cold gas clouds in the MW nuclear wind. (a) ISM clouds in chemical equilibrium are entrained in the hot wind; (b) outer \hi\ envelope is depleted and molecular gas exposed; (c) molecular gas dissociates into \hi; (d) only \hi\ gas left in the cloud; (e) \hi\ ionizes into H\textsc{ii}.
    }
    \label{fig:windcartoon}
\end{figure}

\subsection{A tentative interpretation}

Finally, in this section we propose a speculative interpretation that synthesizes all the observational evidence gathered so far, and is supported by detailed astrochemical modelling (Noon et al., in prep.).
In this picture, illustrated in \autoref{fig:windcartoon}, a hot wind is generated by either the star-formation activity or by supermassive black hole activity in the GC. 
As the hot wind impacts the surrounding interstellar medium, some cold gas clouds become entrained within it. 
Initially, these clouds are in chemical equilibrium and consist of molecular gas clumps embedded within an atomic hydrogen envelope (a). 

As the clouds flow away, ram pressure by the fast-moving hot gas and turbulent mixing at the cloud interface strip the outer \hi\ layer, which is then ionized by the ambient radiation field and cosmic rays (b). 
At this stage, the observed clouds exhibit high molecular gas fractions (Sect.~\ref{sec:molfrac}) and appear to be out of chemical equilibrium, showing an overabundance of H$_2$ relative to \hi\ \citep{Noon+2023}. 
The interaction with the hot wind also increases the level of turbulence \citep{Gerrard+2024}, accounting for the observed broad line widths (Sect.~\ref{sec:veldisp}) and large virial parameters \citep[][and Sect.~\ref{sec:virialpar}]{Heyer+2025}. 
Without the outer \hi\ shielding layer, and under the influence of internal turbulent pressure, the H$_2$ clumps expand and become exposed to the ambient radiation field, which leads to their rapid dissociation and to the reformation of \hi\ (c). 
Once the dissociation process is complete, at distances $\gtrsim 1$~kpc, the remaining clouds are dominated by atomic gas (d) with little or no molecular component (Sect.~\ref{sec:molfrac}). 
At even larger distances, the \hi\ clouds may themselves become ionized or be destroyed by mixing with the surrounding fast-moving hot gas (e) (Lockman et al., in prep.), contributing to the warm ionized gas component detected at higher latitudes with UV absorption studies \citep{Fox+15,Bordoloi+17}.

\section{Summary and conclusions}
\label{sec:conclusion}
In this work, we conducted a systematic search for molecular gas in the nuclear wind of the Milky Way. 
Using the Atacama Pathfinder EXperiment (APEX), we mapped the $^{12}$CO(2$\rightarrow$1) emission line toward \cloudnum\ high–velocity \hi\ clouds that are believed to be entrained in the Galactic Centre outflow.
The data reach a typical sensitivity of $30-80$ mK in a 0.5 $\kms$ velocity channel and have a spatial resolution of $28.6''$, corresponding to $\sim 1$ pc at the distance of the Galactic Centre. 

The main results of our analysis can be summarized as follows.

\begin{itemize}[itemsep=0.2cm]
    \item High–velocity \co\ emission is detected in 12 of the 19 APEX fields. Accounting for secondary high–velocity components that fall within some maps, this yields CO associated with a total of 16 \ion{H}{i} wind clouds. No cloud was detected in \cof\ transition. We also report non-detections in HCN(2$\rightarrow$1) and HCO$^+$(2$\rightarrow$1) for MW-C1.

    \item After segmentation, 207 distinct molecular clumps are identified in the 16 \hi\ clouds with CO detections. 
    Most of \hi\ clouds show a few, well defined molecular clumps each. 
    Three clouds have more complex structures, both in their morphology and kinematics.
    These clumps represent the coldest phase of the known multiphase outflow.

    \item The identified CO clumps are compact, with typical radii $R\sim 1-3$ pc, show relatively large linewidths ($\sigv\lesssim6~\kms$), and have low CO luminosities ($L_{\rm CO}\sim0.3-30$ $~\mathrm{K\,km\,s^{-1}\,pc^2}$) compared to giant molecular clouds in the Milky Way's disk.

    \item We constrained the CO–to–H$_2$ conversion factor in the wind with grids of DESPOTIC radiative–transfer models tailored to the measured $W_{\rm CO}$, clump radii, \hi\ column densities, and turbulent line widths. 
    The inferred $X_{\rm CO(2-1)}$ values range between $5 - 20 \times 10^{20}$ $\xcoun$, several times larger than the canonical Galactic value for the (2$\rightarrow$1) transition.

    \item Wind clumps lie on the classical mass-radius relation for disk molecular clouds, extending it to the low-mass end. 
    Conversely, in the linewidth-radius and linewidth-mass planes, they show systematically larger dispersions than disk clouds, indicating an excess of turbulent motions.
    A virial analysis shows that most clumps have $\mvir\gg\mmol$, with $\avir$ ranging from a few to several hundred, implying that the vast majority ($\sim$90\%) are gravitationally unbound.
    
    \item High velocity dispersions suggest a high level of turbulence, which might be driven by the cloud-wind interaction at the cloud interface, e.g.\ through shear instabilities. 
    Turbulence leads to large internal (turbulent) pressures, $P_\mathrm{int} / k \sim 10^4 - 10^7$ K cm$^3$.
    External confinement by ambient pressure can help some clumps survive, but matching the internal pressures of the full sample would require hot–phase conditions ($n\sim10^{-1}\,\mathrm{cm^{-3}}$, $T\sim10^7$ K) out to $\sim$1 kpc that are unlikely given theoretical expectations and X–ray constraints. A good fraction of clumps is probably expanding, which may lead to the eventual disruption of molecules. 
    
    \item For clouds with CO detections we measure high molecular mass fractions, typically $\fmol\simeq0.6-0.9$, with only two cases near $\fmol\simeq0.3$. 
    We find no clear trend of $\fmol$ with height over $z\lesssim1$ kpc, nor a correlation with peak \hi\ column density. Interestingly, we detect no CO at latitudes $|b|\gtrsim7^\circ$ (projected $z\sim1$ kpc), indicating that molecular gas is rapidly dissociated beyond this height, with typical timescales $\lesssim$ 3 Myr.
    The rapid decline suggests that molecular gas is short–lived in the wind, favoring an entrainment–plus–dissociation scenario rather than \textit{in situ} molecule formation at high height.

\end{itemize}

\noindent In conclusion, our study significantly expands the known molecular component of the Galactic centre wind, providing the first statistical characterization of molecular clouds in the outflow.
Our observations demonstrate that the Galactic centre wind hosts a populous, molecule–rich clump population below $\sim$1 kpc that is kinematically hot, largely unbound, and likely shaped by the interaction with the surrounding hot flow. 
These results highlight that the Milky Way’s nuclear wind is a powerful laboratory for studying cold–hot coupling in multiphase outflows, providing quantitative benchmarks for high–resolution simulations of cloud survival, confinement, and dissociation in galactic winds.

\begin{acknowledgements}
The authors warmly thank the technicians and the scientists who carried out the observations at the APEX telescope.
This work was performed in part at the Aspen Center for Physics, which is supported by National Science Foundation (NSF) grant PHY-2210452, and in part at the Research School of Astronomy and Astrophysics, the Australian National University. 
EDT, AF, NP and QY were supported by the European Research Council (ERC) under grant agreement no.\ 101040751.
MRK and NM-G acknowledge support from the Australian Research Council through Laureate Fellowships FL220100020 and FL210100039, respectively. LA acknowledges support through
the Program ``Rita Levi Montalcini'' of the Italian MIUR. MPB acknowledges support from the National Radio Astronomy Observatory through the Jansky Fellowship.
This publication is based on data acquired with the Atacama Pathfinder Experiment (APEX), under project codes 0104.B-0106A, 0106.B-0034A and 0106.C-0031A. APEX is a collaboration between the Max-Planck-Institut fur Radioastronomie, the European Southern Observatory, and the Onsala Space Observatory.
The GBT data were obtained under project codes 21A\_240 and 22A\_339.
The Green Bank Telescope is a facility of the National Science Foundation and part of the National Radio Astronomy Observatory. It is operated by Associated universities, Inc., under a cooperative agreement with the NSF.  
\end{acknowledgements}

\bibliographystyle{aa}
\bibliography{biblio}

\begin{appendix}
\section{Moment maps for detected CO clouds}
\label{sec:appA}
\begin{figure*}
    \centering
    \includegraphics[width=0.95\textwidth]{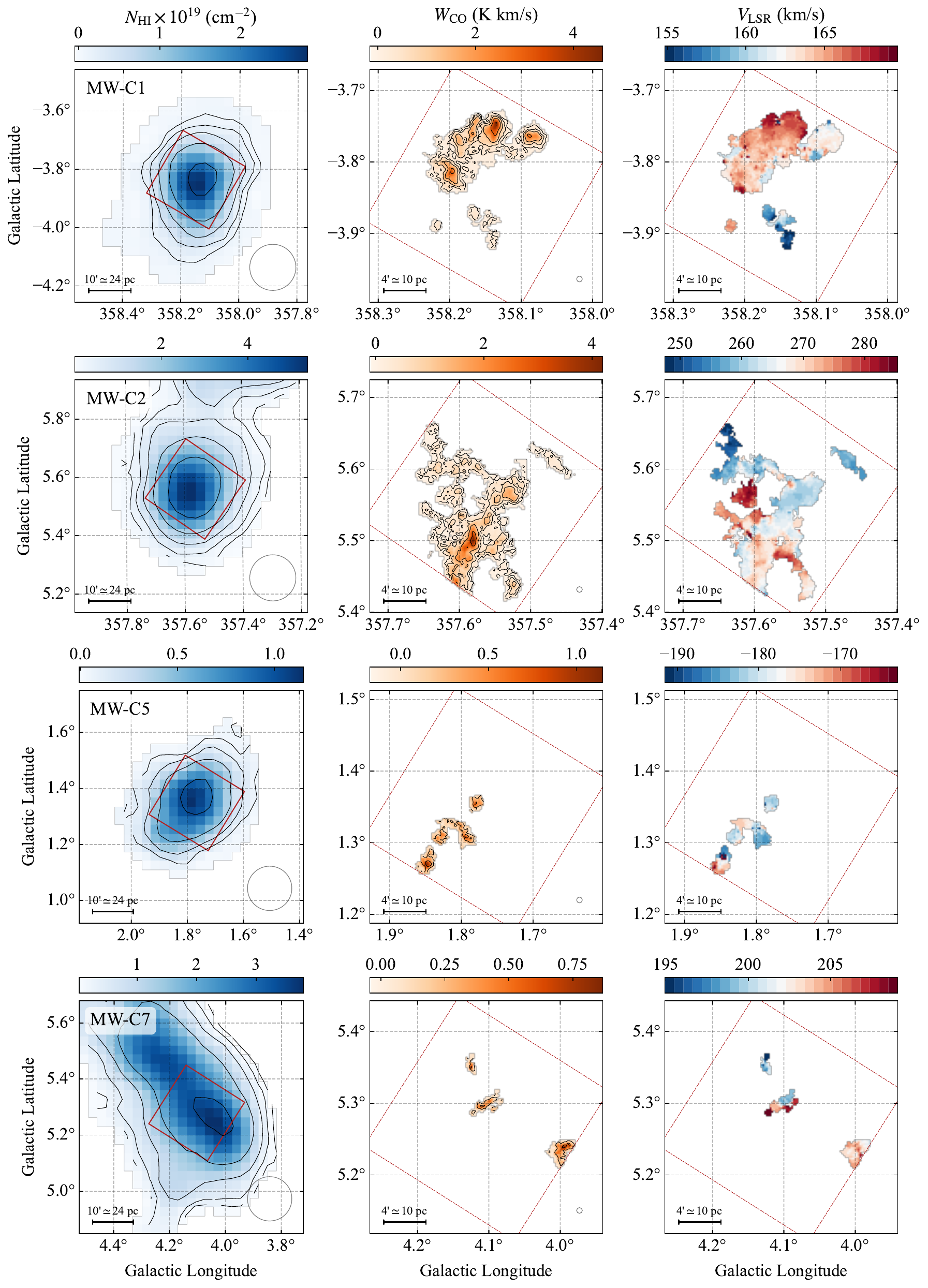}
    \caption{Maps of clouds MW-C1, MW-C2, MW-C5 and MW-C7. See \autoref{sec:appA} for details.}
    \label{fig:maps1}
\end{figure*}

\begin{figure*}
    \centering
    \includegraphics[width=0.95\textwidth]{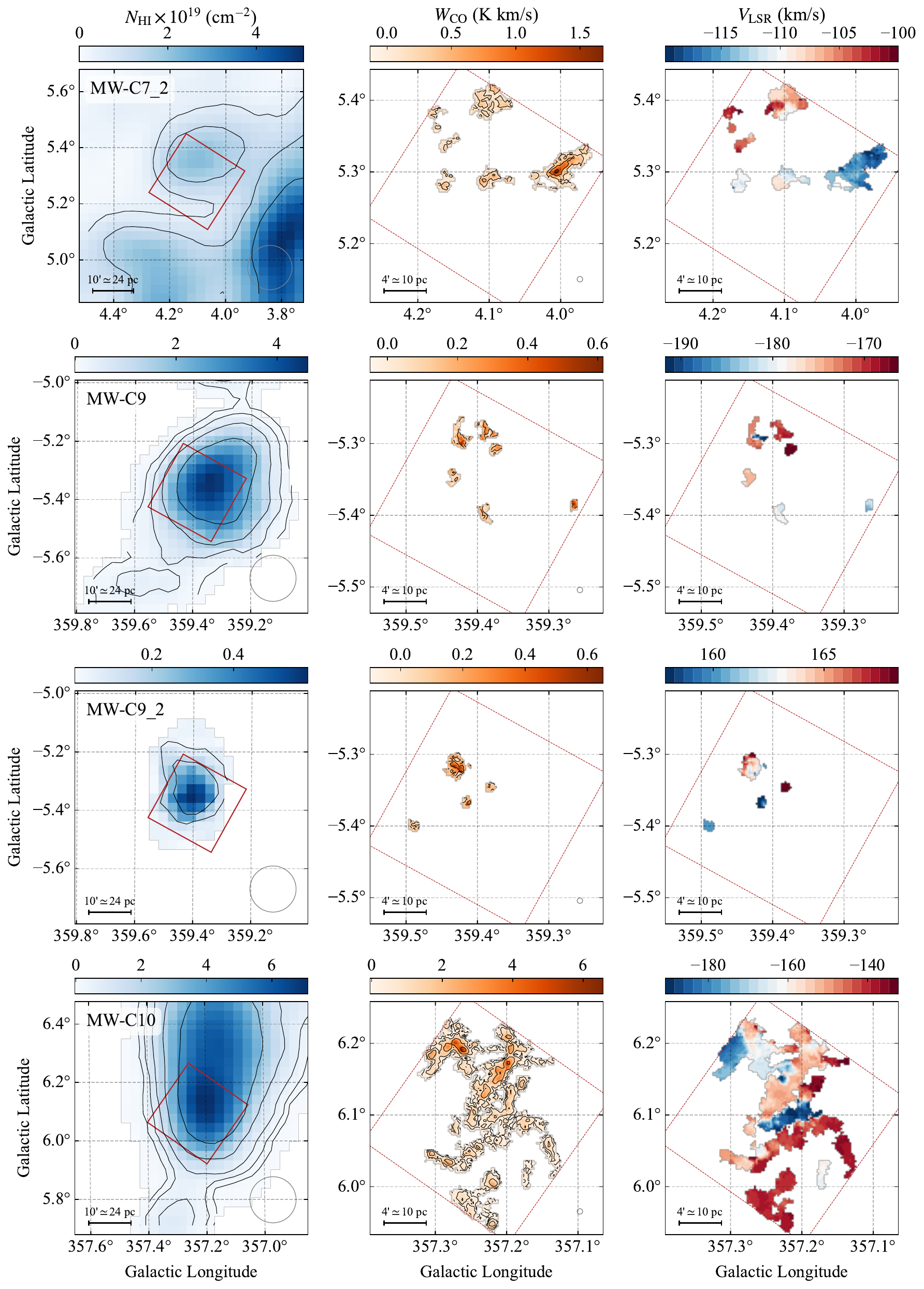}
    \caption{Maps of clouds MW-C7\_2, MW-C9, MW-C9\_2 and MW-C10. See \autoref{sec:appA} for details.}
    \label{fig:maps2}
\end{figure*}

\begin{figure*}
    \centering
    \includegraphics[width=0.95\textwidth]{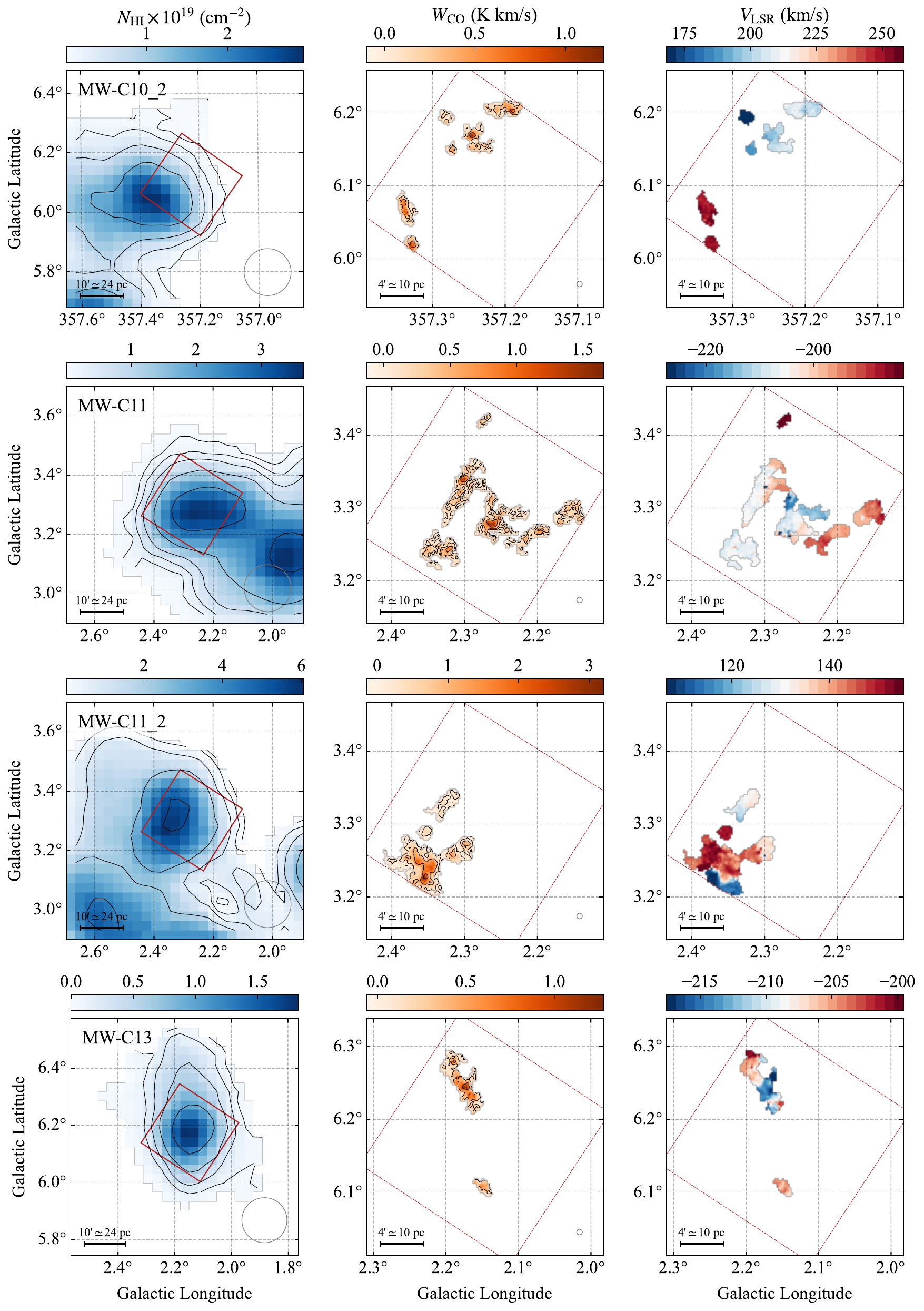}
    \caption{Maps of clouds MW-C10\_2, MW-C11, MW-C11\_2 and MW-C13. See \autoref{sec:appA} for details.}
    \label{fig:maps3}
\end{figure*}

\begin{figure*}
    \centering
    \includegraphics[width=0.95\textwidth]{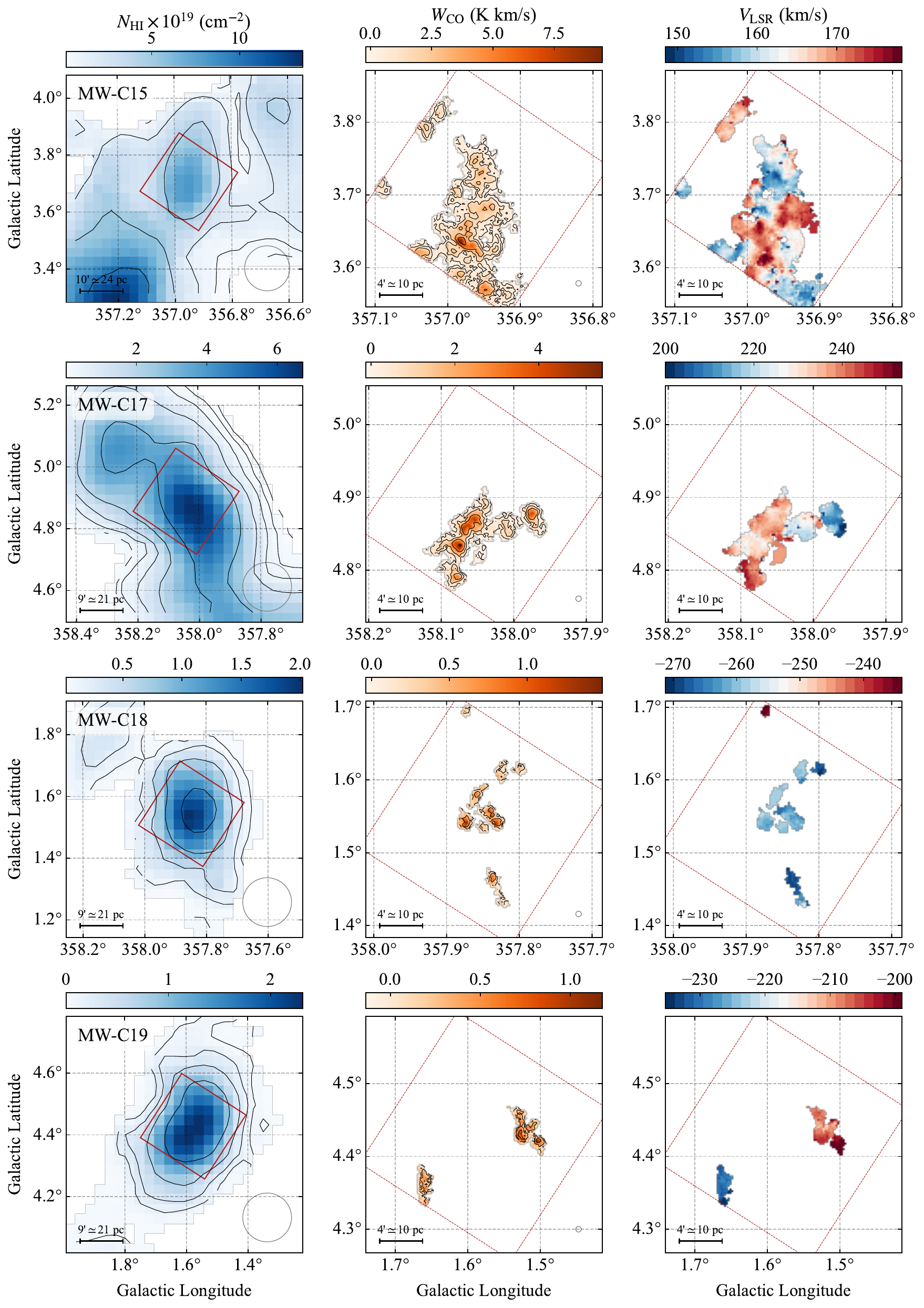}
    \caption{Maps of clouds MW-C15, MW-C17, MW-C18 and MW-C19. See \autoref{sec:appA} for details.}
    \label{fig:maps4}
\end{figure*}

In this Appendix, we show \hi\ and \co\ maps for the 16 clouds with detected molecular gas (Figures~\ref{fig:maps1}-\ref{fig:maps4}).
\hi\ maps are derived from GBT data at $540''$ resolution \citep{DiT+18}, while CO maps are derived from APEX data at $28''$ resolution. 
Each row of panels indicates one \hi\ cloud. 
Clouds labeled with a $\_2$ represent a secondary velocity component detected in the CO APEX data in the same field of the main target. 

From the left to the right, we plot the \hi\ column density map (blue colorscale), the CO integrated intensity map (0th moment, orange colorscale) and the CO LSR velocity map (1st moment, spectral colorscale). 
The red square in all maps highlights the field observed in CO with APEX.
The beam sizes of the GBT and APEX are indicated as grey circles in the \hi\ and CO total maps, respectively.
Contour levels on the \hi\ and CO maps are at $2^n \times 2.5\sigma_\mathrm{rms}$ with $n=0,1,2,3,4$, where $\sigma_\mathrm{rms}$ is the rms noise of the masked 0th moment map \citep[see appendixes of][]{Verheijen+2001,Lelli+2014}.

\section{Table of positions and derived masses}
\label{sec:appB}
\autoref{tab:mass} lists some global properties of the high-velocity clouds in the Milky Way's nuclear wind followed up in molecular tracers to date. 
The table includes the 19 clouds observed with APEX and presented here and in \citet{DiTeodoro+2020} (clouds 1-19) plus the two clouds observed with LMT \citep[clouds 20 and 21][]{Heyer+2025} and reanalysed for this work. 

All quantities are derived assuming a fixed distance of 8.2 kpc from the Galactic Center. Columns as follow:

\begin{enumerate}
    \item Cloud assigned name.
    \item Galactic longitude $\ell$, in degrees.
    \item Galactic latitude $b$, in degrees.
    \item Central local standard-of-rest velocity $\vlrs$, in $\kms$.
    \item Height $z$ above (positive) or below (negative) the Galactic plane, in kpc.
    \item Projected distance $d$ from the Galactic centre, in kpc.
    \item Mass of atomic gas $M_\mathrm{at}$, including Helium, derived from the GBT \hi\ data, in $\mo$.
    \item Adopted molecular gas conversion factors $\alpha_\mathrm{CO}$ (see Section~\ref{sec:radmod}), in $\acoun$. For non-detections, we use the median value across the sample, $\tilde{\alpha}_\mathrm{CO} = 22$  $\acoun$, to estimate upper limits on the molecular mass.
    \item Mass of molecular gas $\mmol$, including Helium, derived from the \co\ APEX data or from the $^{12}$CO(1$\rightarrow$0) LMT  data, in $\mo$. 
    For non-detections, we indicate our upper limits based on the rms noise of the data (see Section~\ref{sec:masses}).
    \item Molecular gas fraction $\fmol = \mmol / (M_\mathrm{at} + \mmol)$. For non-detections, we indicate upper limits.
\end{enumerate}

\begin{table*} \centering \caption{Derived masses of the cloud sample. See \autoref{sec:appB} for details on the columns.} \label{tab:mass} \renewcommand{\arraystretch}{1.3} \begin{tabular}{lccccccccc} \noalign{\vspace{5pt}}\hline\hline\noalign{\vspace{5pt}} Name & $\ell$ & $b$ & \vlrs & $z$ & $d$ & $M_\mathrm{at}$ & $\alpha_\mathrm{CO}$ & $\mmol$ & $\fmol$ \vspace{2pt}\\ & ($\de$) & ($\de$) & ($\kms$) & (kpc) & (kpc) & ($\mo$) & & ($\mo$) & \\ \noalign{\vspace{5pt}}\hline\hline\noalign{\vspace{5pt}} 
MW-C1 & $-1.85$ & $-3.83$ & $+163$ & $-0.55$ & 0.61 & 250 & $23^{+39}_{-13}$ & $3022^{+2369}_{-801}$ & $0.93^{+0.02}_{-0.04}$ \\ 
MW-C2 & $-2.44$ & $+5.56$ & $+265$ & $+0.80$ & 0.87 & 585 & $21^{+45}_{-14}$ & $3792^{+2692}_{-816}$ & $0.87^{+0.03}_{-0.06}$ \\ 
MW-C3$^\dagger$ & $+13.30$ & $+7.13$ & $-100$ & $+1.02$ & 2.16 & 944 & $-$ & $<30$ & $<0.03$ \\ 
MW-C4$^\dagger$ & $+1.66$ & $+5.11$ & $-262$ & $+0.73$ & 0.77 & 105 & $-$ & $<38$ & $<0.27$ \\ 
MW-C5 & $+1.77$ & $+1.35$ & $-178$ & $+0.19$ & 0.32 & 110 & $33^{+35}_{-15}$ & $384^{+178}_{-70}$ & $0.78^{+0.04}_{-0.06}$ \\ 
MW-C6$^\dagger$ & $+8.82$ & $+9.67$ & $+200$ & $+1.38$ & 1.87 & 101 & $-$ & $<20$ & $<0.17$ \\ 
MW-C7 & $+4.10$ & $+5.28$ & $+205$ & $+0.76$ & 0.96 & 560 & $28^{+40}_{-15}$ & $231^{+205}_{-74}$ & $0.30^{+0.09}_{-0.10}$ \\ 
MW-C7\_2 & $+4.14$ & $+5.36$ & $-110$ & $+0.76$ & 0.96 & 265 & $31^{+37}_{-15}$ & $852^{+570}_{-230}$ & $0.77^{+0.06}_{-0.10}$ \\ 
MW-C8$^\dagger$ & $+3.50$ & $+8.62$ & $+365$ & $+1.24$ & 1.33 & 132 & $-$ & $<24$ & $<0.16$ \\ 
MW-C9 & $-0.61$ & $-5.37$ & $-175$ & $-0.77$ & 0.77 & 532 & $36^{+46}_{-16}$ & $238^{+126}_{-45}$ & $0.31^{+0.06}_{-0.07}$ \\ 
MW-C9\_2 & $-0.56$ & $-5.30$ & $+163$ & $-0.77$ & 0.77 & 30 & $52^{+73}_{-24}$ & $242^{+251}_{-83}$ & $0.89^{+0.04}_{-0.07}$ \\ 
MW-C10 & $-2.77$ & $+6.10$ & $-165$ & $+0.87$ & 0.96 & 884 & $20^{+48}_{-14}$ & $6105^{+3217}_{-943}$ & $0.87^{+0.03}_{-0.04}$ \\ 
MW-C10\_2 & $-2.67$ & $+6.04$ & $+210$ & $+0.87$ & 0.96 & 157 & $33^{+41}_{-15}$ & $540^{+266}_{-99}$ & $0.78^{+0.04}_{-0.07}$ \\ 
MW-C11 & $+2.24$ & $+3.28$ & $-203$ & $+0.47$ & 0.57 & 441 & $42^{+66}_{-24}$ & $1543^{+957}_{-350}$ & $0.78^{+0.05}_{-0.08}$ \\ 
MW-C11\_2 & $+2.33$ & $+3.33$ & $+130$ & $+0.47$ & 0.57 & 644 & $17^{+28}_{-10}$ & $1042^{+614}_{-211}$ & $0.62^{+0.07}_{-0.10}$ \\ 
MW-C12$^\dagger$& $-4.75$ & $+8.69$ & $-188$ & $+1.24$ & 1.42 & 151 & $-$ & $<26$ & $<0.15$ \\ 
MW-C13 & $+2.15$ & $+6.18$ & $-217$ & $+0.88$ & 0.94 & 158 & $23^{+33}_{-12}$ & $345^{+264}_{-98}$ & $0.69^{+0.07}_{-0.12}$ \\ 
MW-C14$^\dagger$& $+13.01$ & $-7.64$ & $-182$ & $-1.09$ & 2.15 & 170 & $-$ & $<22$ & $<0.12$ \\ 
MW-C15 & $-3.05$ & $+3.71$ & $+168$ & $+0.53$ & 0.69 & 1015 & $12^{+24}_{-7}$ & $4595^{+2203}_{-703}$ & $0.82^{+0.03}_{-0.05}$ \\ 
MW-C16$^\dagger$& $-1.99$ & $+12.53$ & $-136$ & $+1.79$ & 1.82 & 771 & $-$ & $<32$ & $<0.04$ \\ 
MW-C17 & $-1.96$ & $+4.89$ & $+230$ & $+0.70$ & 0.75 & 918 & $12^{+23}_{-7}$ & $1706^{+1249}_{-411}$ & $0.66^{+0.08}_{-0.11}$ \\ 
MW-C18 & $-2.15$ & $+1.55$ & $-258$ & $+0.22$ & 0.38 & 197 & $37^{+61}_{-20}$ & $769^{+477}_{-160}$ & $0.80^{+0.05}_{-0.07}$ \\ 
MW-C19 & $+1.58$ & $+4.43$ & $-226$ & $+0.63$ & 0.67 & 255 & $43^{+65}_{-22}$ & $553^{+437}_{-149}$ & $0.69^{+0.07}_{-0.11}$ \\ 
MW-C20$^{\star\, \dagger}$ & $+1.55$ & $+3.74$ & $-235$ & $0.53$ & 0.58 & 284 & $-$ & $<22$ & $<0.07$ \\ 
MW-C21$^\star$ & $-1.14$ & $+3.73$ & $+165$ & $0.54$ & 0.56 & 462 & $9^{+16}_{-5}$ & $215^{+216}_{-70}$ & $0.33^{+0.10}_{-0.11}$ \\ \noalign{\vspace{3pt}}\hline \noalign{\vspace*{3pt}} \multicolumn{3}{l}{$^\dagger$ Non-detection.} \\ \multicolumn{5}{l}{$^\star$ $^{12}$CO(1$\rightarrow$0) data from \citet{Heyer+2025}.} \\ 
\end{tabular} 
\end{table*}

\end{appendix}

\end{document}